\documentclass[preprint,12pt,authoryear]{elsarticle}
\usepackage{amssymb}
\usepackage{amsthm}
\usepackage{amsmath}
\usepackage{xcolor}
\usepackage{multirow}
\usepackage{ulem}

\definecolor{greenish}{RGB}{0,128,0}
\definecolor{mygray}{RGB}{180,180,180}
\definecolor{trans_purple}{RGB}{226, 146, 218}
\definecolor{trans_purple_2}{RGB}{218, 183, 230}
\definecolor{trans_red}{RGB}{235,123,123}
\definecolor{trans_blue}{RGB}{102,128,212}

\journal{International Journal of Sediment Research}

\begin{document}

\begin{frontmatter}
\title{On the parameters of common settling velocity models for porous sediment aggregates}

\author[TUD,TUB]{Alexander Metelkin\corref{cor}}
\ead{alexander.metelkin@tu-dresden.de}
\author[TUD,TUB]{Bernhard Vowinckel}
\affiliation[TUD]{organization={Institute of Urban and Industrial Water Management, Technische Universität Dresden}, addressline={Bergstraße 66}, city={Dresden}, postcode={01069}, country={Germany}}

\affiliation[TUB]{organization={Leichtweiß-Institut for Hydraulic Engineering and Water Resources, Technische Universität Braunschweig},
            addressline={Beethovenstraße 51A}, 
            city={Braunschweig},
            postcode={38106},
            country={Germany}}

\begin{abstract}
The settling behavior of sediment aggregates is a critical factor influencing the transport of fine-grained sediments in riverine and marine environments. Due to the small size and fragile structure of cohesive sediment aggregates, direct measurement of their porosity and permeability is challenging. While porosity is often estimated using settling velocity relations, permeability is frequently overlooked. This study examines the impact of considering non-negligible permeability on the properties of flocs. We compare aggregate properties by calibrating experimental data to two settling models in a dilute regime: one assumes a fractal structure of aggregates and neglects permeability, while the other assumes constant porosity and permeability. Our results demonstrate that both models describe the experimental data of highly porous aggregates with similar accuracy. We further investigate aggregate dynamics in more complex flow conditions using numerical simulations employing a volume penalization method to geometrically resolve flocs. We compare the behavior of permeable and impermeable flocs in dense suspension regimes and during dilute settling in density-stratified environments. Our findings reveal that permeability significantly influences settling dynamics in complex scenarios and should be considered when determining aggregate properties.
\end{abstract}


\end{frontmatter}

\newpage
\textcolor{black}{
\textbf{Nomenclature table}
\begin{table}[h!]
    \centering
    \resizebox{\textwidth}{!}{%
    \begin{tabular}{rl}
        \\
        $D_p$ & Diameter of primary particles \\
        $n_f$ & Fractal dimension \\
        $D$ & diameter of coarse grain particle or floc \\
        $C_d$ & Drag coefficient \\
        $C_d^s$, $C_d^{agg}$ & Drag coefficients of sphere and aggregate\\
        $\rho_f$ & Density of fluid \\
        $u$ & Bulk velocity of ambient fluid \\
        $Re$ & Reynolds number \\
        $u_s$ & Settling velocity \\
        $\nu$ & Kinematic viscosity \\
        $g$ & Gravitational acceleration \\
        $\rho_s$ & Density of solid \\
        $R_s$ & Submerged Specific Gravity (SSG) of solid phase\\
        $R_{agg}$ & Submerged Specific Gravity (SSG) of aggregates\\
        $\rho_f$ & Density of fluid \\
        $C_1$, $C_2$ & Laminar and turbulent drag force coefficient \\
        $b_1$,$b_2$ & Generalized laminar and turbulent drag force coefficients\\
        $N$ & Number of primary particles\\
        $\epsilon$ & Porosity \\
        $\kappa$ & Permeability\\
        $\Omega$ & Drag reduction factor\\
        $\beta$ & Non-dimensional permeability\\
        $\alpha$ & Shape factor\\
        $\Phi$ & Sphericity\\
        $\mathcal{A},\mathcal{B},\mathcal{C},\mathcal{D}$ & Polynomial coefficients\\
        $\phi$ & Volume fraction of aggregates in suspension\\
        $\vec{u}$ & Velocity vector of Eulerian phase\\
        $p$ & Pressure of Eulerian phase\\
        $\vec{U}$ & Velocity vector of the aggregates\\
        $t$ & Time\\
        $m_{dry}$ & Dry mass of an aggregate\\
        $m$ & Total mass of an aggregate\\
        $\vec{F_h}$,$\vec{F_g}$,$\vec{F_c}$ & Hydrodynamic, gravity and contact forces acting on aggregates\\
        $\vec{\omega}$ & Angular velocity of an aggregate\\
        $\vec{T_h}$, $\vec{T_c}$ & Hydrodynamic and contact torques of aggregates\\
        $V_{agg}$ & Volume of aggregate\\
        $u_s^{St}$ & Stokes settling velocity\\
        $\rho_b$, $\rho_u$ & Values of density for lower and upper layers in density-stratified environment\\
        $c$ & Non-dimensional concentration field\\
        $D_c$, $D_f$ & Diffusion coefficient in pure liquid and inside porous aggregates\\
        $Pe$ & Peclet number\\
        $\eta$ & Density difference parameter between layers in density-stratified environment\\
        $\zeta$ & Density difference parameter between aggregate and fluid in density-stratified environment\\
        $DST$ & Delayed settling time\\
        \\
    \end{tabular}}
    \label{tab:nomenclature}
\end{table}
}

\newpage
\section{Introduction}
\label{sec:intro}
The fate and transport of sediments plays a key role in various hydrological and environmental processes. It affects morphological properties of estuaries, deltas, rivers, and coasts \citep{dyer1989sediment,seminara2001river, mcanally2001coastal} and has to be considered when studying the carbon cycle for marine ecosystems being both - a vital source of organics and minerals \citep{alldredge1988characteristics} and contamination matter \citep{droppo2001rethinking,grabowski2011erodibility, gillard2019}. 

The dominant part of sediment transport in open water bodies is associated with currents that impose shear on the sediment bed and drag particles along with the flow.  In sufficiently strong flows, sediments are eroded from the river bed and resuspended into the outer flow under the influence of shear and turbulence \citep{vercruysse2017suspended}. The turbulent motion characterizes the mixing of sediments and also affects the internal structure of sediment agglomerates. On the other hand, sediment particles also tend to settle as far as their density is larger than the density of the ambient fluid. Therefore, the inertial properties of sediment grains together with resuspension and deposition mechanisms play a key role in sediment transport \citep{manning2004development,ferguson2004simple}.

To analyze the physics and create large-scale predictions that accurately capture the mechanisms of sediment transport, it is necessary to estimate the density, shape, and, hence, drag coefficient of sediment particles \citep{ferguson2004simple,strom2011explicit}. Consequently, the measurement of sediment particle size, their material, and settling velocity are the main objectives of many experimental campaigns that are used to parameterize sediment transport models \citep{spearman2020measurement,french2010critical}. Based on their size and material, sediments are divided into two main categories - coarse and fine-grained sediments. 
On the one hand, coarse-grained sediments are associated with sands and gravels. Their material properties and settling velocity can be precisely described by existing settling velocity models \citep{ferguson2004simple}, which is why they are out of the scope of the present study. Fine-grained sediments, on the other hand, are particles whose grain size does not exceed 63 $\mu$m and are mostly associated with silt and clay  \citep{grabowski2011erodibility}. Due to their smaller size, fine-grained sediments are affected by cohesive or van der Waals forces that dominate over gravity for this spatial scale \citep{vowinckel2019settling}. Under these forces, platelets of clay flocculate into aggregates called flocculi. Flocculi cannot be broken into smaller units under the local shear conditions and form primary particles of a certain size distribution \citep{winterwerp2004introduction}. These primary particles can further aggregate into bigger structures resulting in the following four-level aggregation chain - \textit{clay platelets - flocculi - flocs - floc aggregates} \citep{leussen1988aggregation}.
Marine snow in the ocean, a vital element in the global carbon cycle, is also frequently associated with flocs and floc aggregates because of the similarities in the material of primary particles and the porous structure \citep{alldredge1988characteristics}. Within the present study, we combine the terms \textit{flocs} and \textit{floc aggregates} and refer to them as aggregates.

Observations show that most of the sediment mass within cohesive suspensions is accumulated in such aggregates \citep{kranck1992characteristics}, which settle faster than their individual primary particles would \citep{manning2006variability}. The internal composition of aggregates and their high porosity obstruct estimations of their density. Aggregates are small, fragile, and hard to extract without damaging their internal structure, which makes it impractical to measure their density directly \citep{eisma1997situ,manning2006variability}. To this end, an approach that estimates the density of a falling aggregate by measuring its settling velocity in a water column is usually applied. The aggregate density is then estimated using the drag coefficient of an equivalent sphere, yet its permeable properties are often overlooked. These assumptions, however, may bear uncertainties when dealing with highly porous aggregates, because permeability affects their total drag force. 

An additional level of complexity is associated with the shape and surface roughness of porous aggregates. The morphological properties of aggregate shapes are usually estimated via  image analysis taken by high-resolution microscopy. Using these methods, it is possible to capture a 2D projection of the aggregate geometry, which is sufficient to estimate its size and the main geometrical properties. Nevertheless, the aggregate then has to be simplified to a spherical object by fitting the value of the captured surface area to a corresponding sphere. This is problematic because there is no single universally accepted method for determining the size of such a sphere \citep{maggi2005}. 

Based on the four-level aggregation chain outlined above, another common assumption is that aggregates have a self-similar, or fractal internal structure. Self-similar properties of aggregates were initially proposed by \citet{krone1963study}, who introduced the concept of orders of fragmentation. These ideas were further advanced by \citet{kranenburg1994fractal}  who introduced a theoretical framework that describes the density of fractal aggregates as a function of their size, the size of their primary particles $D_p$, and a fractal dimension $n_f$ which is a measure for the aggregate compactness \citep{jiang1991fractal}. Both parameters, $D_p$ and $n_f$, depend on the constitutive material and aggregation conditions, but are as difficult to measure directly as permeability. Although assumptions about the fractal geometry of natural aggregates are widely used, recent research suggests that they might not be appropriate to describe porous aggregates. By using a novel sampling technique utilizing 3D nanotomography to capture the internal composition of the natural flocs, \citet{spencer2021structure} have shown that the structure of aggregates is not self-similar and, therefore, cannot be described in terms of fractal theory. Hence, the widely accepted assumptions of aggregate fractal nature should be reconsidered, and the dependence of aggregate properties on its size should be redefined.

The huge variability of parameters that are considered in settling velocity models together with the high dispersion of the measurement data impede the exact determination of the governing mechanisms responsible for the modification of the settling velocity. In addition, there are no accepted ranges of parameters that could be helpful to estimate the real properties of sediments. This is partly due to the fact that  parameters such as fractal dimension, aggregate shape, or the drag reduction caused by permeability are generally not constant values and depend on the aggregate size \citep{maggi2005,maggi2007variable,van1994estuarine,vanni2000creeping}. Nevertheless,  they are usually treated as constant for the sake of simplification.

In this regard, \citet{strom2011explicit} derived one of the most comprehensive settling velocity models that is capable of incorporating a wide range of \textit{in-situ} and laboratory measurements. The model is based on the relation proposed by \citet{ferguson2004simple}, which includes the coefficients for viscous and turbulent drag in a single equation and combines fractal-based density relations together with coefficients that are associated with the shape and permeability of aggregates. Such a model is capable of predicting the settling velocity of individual aggregates, but it contains many parameters that are tunable because they are usually unknown. The authors also analyzed models that accounted for a variable fractal dimension of aggregates, but the effects of the permeability and shape of aggregates are represented in a rather simplified manner because they are expressed within a single constant parameter only. While these measures allow for convenient parameterization of individual settling aggregates by calibration against experimental data, the origin of drag force modification of aggregate cannot be designated.

The discussion laid out so far has shown that, on the one hand, aggregate properties such as permeability, shape, and density are important but very hard to measure directly with high accuracy using conventional experimental methods. On the other hand, however, recent evidence has been presented that fractal assumptions as a surrogate integral quantity to describe aggregate composition may not be an accurate description for natural sediment aggregates, either. This discrepancy is addressed in the present study. The settling velocity model introduced by the seminal study of \citet{strom2011explicit} appears to capture the experimental data with a high coefficient of determination, but the range of parameters that lead to good predictions is wide, potentially providing ambiguous definitions of the floc properties. This can yield results, where different combinations of fractal dimension and porosity provide good estimates of the settling velocity in an unbounded flow, but provide different hydrodynamic properties in more complex flow configurations.
This may have important consequences on the settling behavior. Determining, let alone even tuning, the hydrodynamic parameters of a floc is extremely challenging in experimental setups \citep{spencer2021structure}. Hence, direct numerical simulations (DNS) offer an alternative way to study more complex flow situations with controlled boundary conditions.
The DNS framework geometrically resolves aggregates as spherical objects and treats the interior as a continuum through the Brinkman-extended Darcy equations for porous objects \citep{chandesris2007boundary,neale1974practical}. Aggregate-resolving simulations therefore allow to precisely tune floc properties, such as fractal dimension, porosity and permeability to compare the settling behavior for different scenarios. Recent studies of particle-resolved simulations of non-porous objects using the Immersed Boundary method have shown promise to obtain highly resolved information on the settling behavior for unbounded settling of individual particles \citep{biegert2017collision} and hindered settling of dense suspension \citep{vowinckel2019settling,vowinckel2019consolidation}. To address the implications of different parameter combinations that describe the hydrodynamic properties of porous particles and govern their settling behavior, we extend the particle-resolving framework along the lines of \citet{panah2017simulations}. These authors investigated the settling of a single porous aggregate in a two-dimensional domain. In this study, we extend these consideration to aggregate-resolving simulations of many porous objects submerged in a viscous fluid in three-dimensional domains, thereby providing a new means to simulate the transport of porous aggregates in viscous flows with high fidelity. This allows to test the implications of different parameter combinations for a variety of flow configurations, such as hindered settling and settling in density stratified flows.

The paper is structured as follows. First, we review common settling velocities in section \ref{sec:models}. Subsequently, we determine appropriate ranges for the parameters shape factor, aggregate density and permeability in section \ref{sec:parameters} to derive two new surrogate models from the considerations of \citet{strom2011explicit} that aim to simplify their model by isolating effects of fractal dimension and permeability, respectively. We then show the ambiguity of the results provided by parameterisation in section \ref{sec:fitting_with_experiments} and demonstrate the implications in section \ref{ch:num_sim}. Finally, section \ref{sec:conclusions} summarizes the main conclusions and provides an outlook for future research.

\section{Common settling velocity models}\label{sec:models}
Settling velocity models are represented by algebraic expressions derived for unbounded conditions and based on the drag relations of a sphere of size $D$ in a uniform flow. For coarse-grained sediments, the settling velocity can be determined for individual grain sizes, whereas for fine-grained sediments, the settling velocity is determined for aggregate sizes comprising many primary particles. Due to the difference in internal structure between individual grains and aggregates, the settling velocity models for these two types of sediments are derived differently.

Within the present section, we first review common settling velocity models for coarse and fine-grained sediments, respectively. Then, we discuss the parameters that affect the settling rate for the fine-grained sediments and define the ranges of tunable variables that are not accessible. These variables will be used further as fitting parameters for calibration. We also introduce settling velocity models that isolate the effects of density reduction due to the fractal nature of aggregates and permeable drag reduction. Finally, we compare all presented settling velocity models with the experimental data by fitting them to corresponding measurements and discuss the difference between the model estimations in terms of settling rate and aggregates' properties.

\subsection{Coarse-grained sediments}
The general formula for the drag force of a sphere in a uniform flow with velocity magnitude $u$ is defined as $F_d = \frac{1}{8} \pi C_d \rho_f u^2  D^2$, where $C_d$ is the drag coefficient, $\rho_f$ is the density of the ambient fluid. In the context of a falling object in an otherwise quiescent fluid, the velocity magnitude of the flow is represented by the settling velocity of the object $u_s$. The Reynolds number is therefore also defined by taking into account this settling velocity $Re = u_s D/\nu$, where $\nu$ is the kinematic viscosity of the ambient fluid. Balancing the drag force with the force of gravity,  the settling velocity of a sediment particle becomes:
\begin{equation}
    \label{eq:Settling_velocity_general}
    u_s = \sqrt{\frac{4R_{s}Dg}{3 C_d}} \qquad,
\end{equation}

where $g$ is the gravitational acceleration, $R_{s} = (\rho_{s}-\rho_f)/\rho_f$ is the Submerged Specific Gravity (SSG) of a sediment particle, and $\rho_{s}$ is the density of the sediment material. For Stokes flow ($Re \ll 1$), the drag coefficient for an ideal sphere is analytically derived as $C_d = 24/Re$ and, hence, Eq. \ref{eq:Settling_velocity_general} simplifies to
\begin{equation}
    \label{eq:Stokes_settling_velocity}
    u_s = \frac{R_{s} D^2g}{C_1 \nu} \qquad ,
\end{equation}
where $C_1 = 18$ is an analytical coefficient that is related to the spherical shape.

For higher Reynolds numbers, there is no analytical solution for $C_d$, and hence, semi-empirical correlations are used to find a drag coefficient as a function of $Re$ \citep{flemmer1986drag}, while simplified models approximate $C_d$ as constants \citep{cheng1997simplified}. Using dimensional analysis, \citet{ferguson2004simple} derived a relation for settling velocity that combines Stokes and turbulent drag models, resulting in the following relation for the settling velocity:
\begin{equation}
    \label{eq:sv_ferg}
    u_s = \frac{R_{s} D^2g}{C_1 \nu + \sqrt{0.75 C_2 R_s D^3 g}} \qquad,
\end{equation}
where $C_2$ is the asymptotic value of the turbulent drag coefficient at $10^3<Re<10^5$. 

In the following sections, we use this settling velocity model as a reference for analyzing experimental measurements of the settling behavior of fine-grained sediments. Hence, within the application to fine-grained sediments the SSG of sediment material $R_s$ is interchanged with the SSG of sediment aggregate $R_{agg}$. In what follows, we use the label \textit{FC} to refer to the model presented by Eq. \ref{eq:sv_ferg}.

\subsection{Fine-grained sediments}
Assuming the self-similar structure of fine-grained aggregates, the aggregate size $D = D_p \cdot N^{1/n_f}$ depends on the total number of flocculi $N$, their diameter $D_p$ and the fractal dimension $n_f$. The relation for the SSG of an aggregate can then be established as  \citep{kranenburg1994fractal}:

 \begin{equation}
    \label{eq:fractal_density_diff}
    R_{agg} = \frac{\rho_{agg}-\rho_f}{\rho_{agg}} = R_s\left(\frac{D}{D_p}\right)^{n_f-3} \qquad ,
\end{equation}
where $\rho_{agg} = \rho_f \epsilon + \rho_s (1-\epsilon)$ is the density of the aggregate with porosity $\epsilon$ and $R_s$ is the submerged specific gravity of the flocculi, which can be assumed to be equal to the SSG of clay material.

By comparison with solid coarse-grained particles, the calculation of the total drag force of a porous aggregate also acknowledges that fluid can pass through its porous structure. The permeable properties of aggregates are defined by the porosity $\epsilon$ and the permeability $\kappa$ that can vary radially and also depend on the fractal dimension and the size of primary particles \citep{vanni2000creeping}.  
For simplicity's sake, we assume homogeneous permeability and porosity inside the porous aggregate. These assumptions allow to define a drag reduction factor $\Omega$ that represents the ratio of the drag force of a permeable sphere to that of a solid sphere as  \citep{brinkman1947calculation,ooms1970frictional,neale1973creeping}
\begin{equation}
    \label{eq:Omega}
    \Omega = \frac{\mathrm{Drag\:of\:permeable\:sphere}}{\mathrm{Drag\:of\:solid\:sphere}} =\frac{2\beta^2(1-\beta^{-1})}{2\beta^2+3(1-\beta^{-1})} \qquad .
\end{equation}
Here, $\beta$ is the non-dimensional permeability that depends on the aggregate diameter and its permeability $\kappa$:
\begin{equation}
    \label{eq:beta}
    \beta = \frac{D}{2\sqrt{\kappa}} \qquad.
\end{equation}
It is not trivial to calculate the value of permeability for a separate aggregate, but there are several empirical models that estimate $\kappa$ considering the primary particle diameter $D_p$ and aggregate porosity $\epsilon$ \citep{vanni2000creeping}. In the present work, we use a common model to estimate permeability as \citep{macdonald1979flow,vanni2000creeping,breugem_boersma_uittenbogaard_2006}

\begin{equation}  
    \label{eq:permeability}
    \kappa = \frac{D_p^2 \epsilon^3}{180(1-\epsilon)^2}  \qquad .
\end{equation}

To account for the effect of the fractal structure and permeability of flocs on the settling velocity, \citet{strom2011explicit} combined equations Eq.\ref{eq:sv_ferg} and Eq.\ref{eq:fractal_density_diff} into a model for settling velocity that comprises the shape, sphericity, and surface roughness of the flocs through the coefficient $\alpha$

\begin{equation}
    \label{eq:sv_strom_orig}
    u_s = \frac{R_s D^{n_f-1}g}{b_1 \nu D_p^{n_f-3} + b_2 \sqrt{ R_s D^{n_f} D_p^{n_f-3}g}} \qquad ,
\end{equation}
where $b_1 = \alpha \Omega C_1$ and $b_2 = \sqrt{0.75 \alpha \Omega C_2}$ . In what follows, we refer to Eq.\ref{eq:sv_strom_orig} as model \textit{SK}. Note, that the value of $\alpha$ does not correlate directly with the sphericity or roughness. Rather, it reflects an empirically defined coefficient that quantifies the deviation of the settling behavior of an ideal sphere from that of an irregularly shaped aggregate.

\section{Aggregate parameters that affect the settling rate}\label{sec:parameters}

As seen from Eq.\ref{eq:sv_strom_orig}, the general formulation of settling velocity consists of many parameters that are nearly impossible to measure directly. This makes it difficult to understand the exact difference between the settling dynamics of flocs and solid sediment particles. In the following, we elaborate on the governing mechanisms affecting the settling velocity by introducing separate settling velocity models. We address the three main mechanisms that affect the settling velocity of porous aggregates: (i) shape, (ii) density, and (iii) permeable properties. Focusing on the latter two, we propose settling models that isolate the effects of density reduction and the influence of permeable drag reduction.

\subsection{Shape}
\label{ch:shape}
\definecolor{myblue}{RGB}{204,204,255}
\definecolor{mygreen}{RGB}{200,240,200}
\begin{figure}[b!]
        \centering
        \includegraphics[]{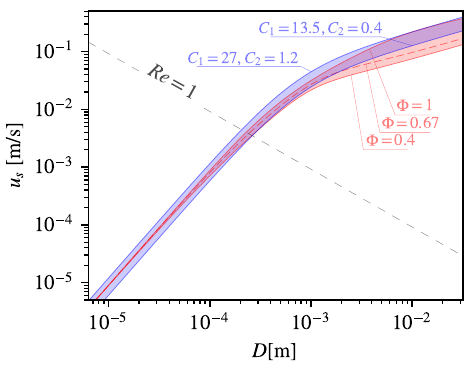}
        \caption{Settling velocity as a function of aggregate diameter. The range \textcolor{pink}{\rule[0.0mm]{0.75cm}{5pt}} indicates solutions of equation Eq.\ref{eq:Cd_sphericity} for various sphericity values ($0.4 < \Phi < 1.0$). $\Phi =1$ corresponds to spherical shapes, $\Phi = 0.67$ correspond to tetrahedrons and $\Phi = 0.4$ to disks \citep{haider1989drag}. The range \textcolor{myblue}{\rule[0.0mm]{0.75cm}{5pt}} corresponds to Eq.\ref{eq:sv_ferg} for minimal and maximal values of Stokes and turbulent drag coefficients. } 
        \label{fig:sphericity_plot}
\end{figure}

The first mechanism that governs the settling velocity of complex aggregates is related to the shape of aggregates. It is important to note that different studies account for the shape and surface roughness of the sediment particle differently. Some studies consider variability in $C_1$ and $C_2$ \citep{cheng1997simplified,ferguson2004simple}, while others keep $C_1$ and $C_2$ constant and account for the shape and roughness via $\alpha$ \citep{strom2011explicit}. Ultimately, both approaches have the same physical meaning with respect to the parameter $\alpha$, since it is directly multiplied by the coefficients $C_1$ and $C_2$ in the corresponding terms of the denominator. For the present study, we choose the former approach by considering the shape factor through the coefficients $C_1$ and $C_2$.  This measure essentially replaces $\alpha$. 

One of the morphological properties of an aggregate that is usually used in the analysis of floc shape is sphericity $\Phi$ as the ratio of the surface of an aggregate and the surface of a sphere of equal volume \citep{haider1989drag}. However, due to the small size of sediment aggregates, it is challenging to estimate the three-dimensional shape. Alternatively, shape irregularity might be considered via circularity - a 2D analog of sphericity that is found as the ratio of a 2D projection perimeter of an aggregate to the perimeter of a sphere that has the same projection area. The study by \citet{haider1989drag} summarizes experimental data sets that contain drag coefficients for various shapes (spheres, cubes, octahedrons, tetrahedrons, and non-isometric bodies) and proposes an equation for the drag coefficient as a function of sphericity  $\Phi$ and Reynolds number $Re$:
\begin{equation}
    \label{eq:Cd_sphericity}
    C_d = \frac{24}{Re} \left( 1+\mathcal{A} Re^\mathcal{B} \right ) + \frac{\mathcal{C}}{1+\frac{\mathcal{D}}{Re}},
\end{equation} 
where the parameters $\mathcal{A}$,$\mathcal{B}$,$\mathcal{C}$ and $\mathcal{D}$ are again polynomial functions of $\Phi$. Using equation Eq.\ref{eq:Settling_velocity_general} with Eq.\ref{eq:Cd_sphericity} for various $\Phi$, one can estimate how the irregularity of the shape influences the settling behavior. Figure \ref{fig:sphericity_plot} shows a range of settling velocities as a function of aggregate diameter for various sphericity values. For this example, SSG is set constant as $R_{agg}=0.15$ for all cases, which corresponds to a density of water $\rho_f= 1046.5$ [$kg/m^3$] and the excessive density of floc as 160 [$kg/m^3$] \citep{manning2006variability,ye2020oil}.
It is seen from the figure \ref{fig:sphericity_plot} that the settling velocity of a spherical object is almost the same as that of an object with a sphericity value of $\Phi = 0.4$ for $Re < 1$. A similar approach introduced by \citet{FRANCALANCI2021118068} considers the variation of the drag coefficients via the Corey Shape Factor ($CSF$). Although the analysis of their model is not present in this paper, it demonstrates that varying the $CSF$ within the range from 1 (spherical particle) to 0.7 results in minimal differences in settling velocities for $Re < 1$. The value of $CSF=0.7$ corresponds to natural sediment grains according to \cite{dietrich1982settling}.

As yet another way to account for shape, it is also possible to simply set drag coefficients  constant within certain limits. For example, \citet{cheng1997simplified} compared the settling velocity of sand grains to spheres of the same size and found that the drag coefficient can increase by up to 42~\%. Similarly, numerical simulation results \citep{dietzel2013numerical,dietzel2016application} show that the drag coefficient of an aggregate $C_d^{agg}$ compared to that of a sphere $C_d^{s}$ varies in the range $0.75<C_d^{agg}/C_d^{s}<1.5$. This range yields corresponding intervals for $C_1=[13.5,27]$ and $C_2=[0.4,1.2]$. Figure \ref{fig:sphericity_plot} shows the settling velocity as a function of aggregate size by applying equation \ref{eq:sv_ferg} and using the corresponding limits for coefficients $C_1$ and $C_2$. The settling velocity is represented by the blue lines, while the semi-transparent filling between these lines contains all possible solutions of the equation \ref{eq:sv_ferg} with the coefficients $C_1$ and $C_2$ within the specified ranges.

It is seen from the figure \ref{fig:sphericity_plot} that these ranges of $C_1$ and $C_2$ cover the variability for the formulation that uses sphericity in the range $\Phi = [0.4,1]$ for $Re<1$. It can be concluded that different parametrizations of the aggregate shape yield comparable results for the particle settling velocity in this viscously dominated regime. Due to this, we consider the ranges of $C_1 = [13.5,27]$ and $C_2 = [0.4,1.2]$ as fitting ranges for the calibration of the settling velocity models to compare against various experimental measurements.

\subsection{Aggregate density}
\label{ch:density}
The second mechanism that affects the settling velocity is related to the assumptions of fractal theory, which defines the density of a floc as the function of its size via Eq.\ref{eq:fractal_density_diff}. The parameters that are related to this mechanism are (1) fractal dimension $n_f$, (2) the mean diameter $D_p$ of primary particles, i.e. the flocculi, and (3) the SSG of the solid material $R_s$. 

The density of primary particles is typically estimated as the density of sediment material and corresponds to that of quartz. This makes $R_s$ a defined parameter. On the contrary, the estimation of flocculi size distribution is usually not available. It is also hardly possible to capture the porosity of an aggregate via image analysis and hence, the fractal dimension of flocs cannot be estimated, either. Consequently, the fractal dimension $n_f$ and the mean diameter of the primary particles $D_p$ are used as calibration parameters for settling velocity models. We prescribe the mean diameter of primary particles to be in the range $1 \, \mu m < D_p < 50 \, \mu m$. This variation in size covers the range from medium-sized clay particles up to the sizes of microflocs ($ \sim 50 \mu m$) \citep{lee2012multimodality}. The fractal dimension range is defined by the geometrical definition of $n_f$ and lies between 1 and 3 \citep{kranenburg1994fractal}. The highest value of fractal dimension $n_f =3$ corresponds to a completely non-porous aggregate with a density equal to the density of clay material, while $n_f$ that limits to 1 is related to extremely high values of aggregate porosity, i.e. $\epsilon \to 1$.

To isolate the mechanism of density reduction according to the fractal structure of aggregates, we simplify the settling velocity model Eq.\ref{eq:sv_strom_orig} by neglecting the effect of permeability by setting $\Omega = 1$:

\begin{equation}
    \label{eq:sv_frac}
    u_s = \frac{R_s D^{n_f-1}g}{C_1 \nu D_p^{n_f-3} + \sqrt{0.75 C_2 R_s D^{n_f} D_p^{n_f-3}g}} \: .
\end{equation}

Eq.\ref{eq:sv_frac} will be referred to as model \textit{FR}. We use the variables $D_p$ and $n_f$ as fitting parameters for further calibration of the model to experimental measurements while keeping $C_1$ and $C_2$ constant.

\subsection{Permeability}

Even though the influence of the permeability is neglected in many studies that use floc settling velocity models \citep{winterwerp2006heuristic,manning2006variability,kumar2010floc}, it is considered in the analysis of settling velocity equations of \citet{strom2011explicit}. These authors acknowledge the dependency of $\Omega$ on aggregate size but in the general formulation of Eq.\ref{eq:sv_strom_orig}, $\Omega$ is simplified to a constant value. In the present study, we drop this assumption and use Eq.\ref{eq:Omega} to capture the dependency of $\Omega$ on aggregate size and permeability. Similarly to Eq.\ref{eq:sv_frac}, we simplify Eq.\ref{eq:sv_strom_orig} in order to suppress the effects of variable density and the shape of an aggregate, which leads to the following settling velocity model:

\begin{equation}
    \label{eq:sv_porous}
    u_s = \frac{R_{agg} D^2g}{ C_1 \nu \Omega + \sqrt{0.75 \Omega C_2 R_{agg} D^3 g}} \qquad .
\end{equation}
For this model, the aggregate submerged specific gravity $R_{agg}$  is treated as an unknown quantity that is used as a fitting parameter together with $D_p$. The diameter of primary particles affects the permeability $\kappa$ and hence $\Omega$ that is calculated via Eq.\ref{eq:Omega}. This eventually defines $\Omega$ as a function of $D$, $D_p$ and $R_{agg}$. We refer to Eq.\ref{eq:sv_porous} as model \textit{PE}.

\section{Results}
\subsection{Applying settling models to experimental data}
\label{sec:fitting_with_experiments}
 
After having established the parameters of the different settling models in section \ref{sec:parameters}, we proceed to evaluate their performance using experimental data on settling velocities for different types of cohesive sediments. We exemplify this by using the experimental data reported by \citet{gillard2019}, who analyzed the settling velocity of the deep-sea sediments that flocculated under different flow conditions. Additional comparisons of the settling velocity models and further experimental data are provided in Appendix B.

The first step of the analysis is to compare the models that have previously been reported in the literature, i.e. \textit{FC} and \textit{SK}. While the model \textit{FC} is widely used for the analysis of various sediments \citep{fan2023interaction,yang2023settling,cuttler2017estimating}, the model \textit{SK} is developed specifically for mud flocs \citep{strom2011explicit}. Both models have free parameters that are usually determined by calibration to optimize the coefficient of determination ($R^2$). Note, that by calibration we imply any technique that minimizes the difference between the model and the experimental data. For the present study, we employ a nonlinear least-squares optimization to find the best-fitting parameters for the settling velocity model. For the \textit{FC} model, the parameters $C_1$, $C_2$, and $\epsilon$ are adjusted, while for the \textit{SK} model, the parameters $C_1$, $C_2$, $D_p$, and $n_f$ are used for fitting. According to our analysis in section \ref{sec:parameters}, we fit the drag coefficients $C_1$ and $C_2$ for both models within the ranges (i.e. $13.5 < C_1 < 27$ and $0.4 < C_2 < 1.2$). For $D_p$ and $n_f$ that affect the \textit{SK} model, we choose $1.0 < n_f < 3.0$ and $1 \ [\mu m] < D_p < 50 \ [\mu m]$, respectively. For the model \textit{FR}, we fit $\epsilon$ in the range $0 < \epsilon < 1$ which in turn affects  $R_{agg} = (\rho_{agg}-\rho_f)/\rho_{agg}$ via $\rho_{agg} = \rho_f \epsilon + (1-\epsilon) \rho_s$.

\begin{figure*}[!t]
     \begin{center}
         \includegraphics[width=8cm]{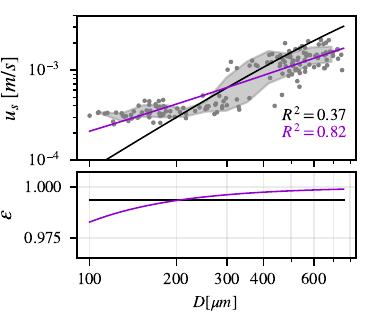}
         \put(-240,170){$a)$}
         \put(-240,80){$b)$}
     \end{center}
     
     \caption{     
     Settling velocity \textit{a)} and porosity \textit{b)} as a function of aggregate size of the \textit{SK} (\textcolor{violet}{\rule[0.5mm]{0.5cm}{1pt}}) and \textit{FR} (\textcolor{black}{\rule[0.5mm]{0.5cm}{1pt}}) models compared with the experimental data set of \citet{gillard2019}. The experimental data (\textcolor{gray}{$\bullet$}) correspond to the sediment concentration of 500 [$mg/l$] and zero shear rate during flocculation. The grey area (\textcolor{mygray}{\rule[0.0mm]{0.5cm}{5pt}}) in (a) represents the normal distribution of the experimental data with a cut-off of the standard deviation $\sigma$.}
\label{fig:SK_FC_comp}
 \end{figure*}

The resulting best-fit curves are shown in figure \ref{fig:SK_FC_comp}a. It can be seen from the plot that the \textit{FC} model (black line) shows a faster increase in settling velocity with increasing aggregate size. Because of this difference in slope, the \textit{FC} model is unable to match the experimental values properly. This is also reflected in the corresponding rather poor value of $R^2=0.37$. On the contrary, the slower increase in settling velocity for the \textit{SK} model yields a much better correlation ($R^2 = 0.82$). The reason for this behavior can be found in the different ways porosity is treated in the two models \textit{FC} and \textit{SK}. This is illustrated in figure \ref{fig:SK_FC_comp}b, where the porosity $\epsilon$ becomes a function of aggregate size via equation \ref{eq:fractal_density_diff} for the \textit{SK} model while it is set constant by definition for the \textit{FC} model. Consequently, the \textit{SK} model is more versatile in capturing various slopes of settling velocity versus aggregate size encountered in the experimental data. The inability of the \textit{FC} to capture the deviation of settling velocity from experimental data is visible for most of the data sets analyzed in figure \ref{fig:SK_FC_comp} and Appendix B. Therefore, we find the \textit{FC} model inadequate to describe the settling behavior of porous aggregates and neglect it for the remainder of the analysis of this section. Instead, we focus on the implications of the parameterization of \textit{SK} model in the next paragraphs.

 \begin{figure*}[!t]
     \begin{center}
         \includegraphics[width=8cm]{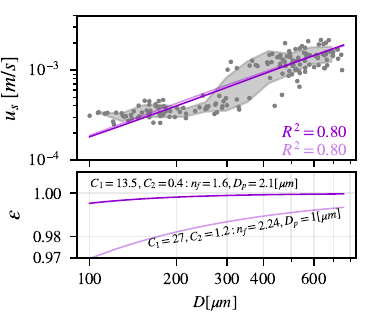}
         \put(-240,170){$a)$}
         \put(-240,80){$b)$}
     \end{center}
     
     \caption{
      Same as figure \ref{fig:SK_FC_comp} but for the \textit{SK} model with different values of $n_f$, $D_p$, $C_1$ and $C_2$. The line \textcolor{violet}{\rule[0.5mm]{0.5cm}{1pt}} corresponds to lowest drag coefficients $C_1 = 13.5$ and $C_2 = 0.4$ resulting in optimal values of $n_f = 1.6$, $D_p = 2.1 [\mu m]$, while the line \textcolor{trans_purple}{\rule[0.5mm]{0.5cm}{1pt}} corresponds to $C_1 = 27$ and $C_2 = 1.2$ resulting in optimal values of $n_f = 2.24$, $D_p = 1 [\mu m]$.}
\label{fig:SK_all_fit}
 \end{figure*}

  \begin{figure*}[!t]
     \begin{center}
         \includegraphics[width=8cm]{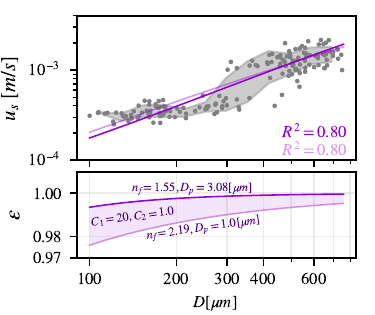}
         \put(-240,170){$a)$}
         \put(-240,80){$b)$}
     \end{center}
     
     \caption{Same as figure \ref{fig:SK_FC_comp}, but for  different values of $n_f$ and $D_p$  of the \textit{SK} model, while $C_1$ and $C_2$ are fixed. The line \textcolor{violet}{\rule[0.5mm]{0.5cm}{1pt}} corresponds to $n_f = 1.55$ and $D_p = 3.08 [\mu m]$, the line \textcolor{trans_purple}{\rule[0.5mm]{0.5cm}{1pt}} corresponds to $n_f = 2.19$ and $D_p = 1 [\mu m]$. The shaded purple area (\textcolor{trans_purple_2}{\rule[0.0mm]{0.5cm}{5pt}}) in (b) represents all possible solutions of the \textit{SK} model that correlate with experimental data with $R^2 > 0.8$.}
\label{fig:SK_C1_fix}
 \end{figure*}

The \textit{SK} model has four degrees of freedom that describe the hydrodynamic properties, where the drag is described by $C_1$ and $C_2$ and the porosity and the permeability are both functions of $n_f$ and $D_p$ (equations Eq.\ref{eq:fractal_density_diff} and Eq.\ref{eq:Omega}). This makes the model flexible on the one hand, but may lead to several parameter combinations that yield similar results on the other hand. To explore this issue, we set the drag coefficients $C_1$ and $C_2$ to their respective bounds (i.e. $C_1=[13.5,27]$ and $C_2=[0.4,1.2]$) and determine the values of $D_p$ and $n_f$ within which the model would yield a high $R^2$. In order to do this, we use the same experimental dataset shown in figure \ref{fig:SK_FC_comp}.The resulting settling velocity curves are shown in figure \ref{fig:SK_all_fit}a and the corresponding porosity values are shown in figure \ref{fig:SK_all_fit}b. Choosing the extreme values for $C_1$ and $C_2$ produces two distinct model parametrizations: (1) setting $C_1 = 27$ and $C_2 = 1.2$  yields $n_f = 2.24$ and $D_p = 1 [\mu m]$, which corresponds to less porous aggregates, and (2) setting  $C_1 = 13.5$ and $C_2 = 0.4$, we obtain $n_f = 1.6$ and $D_p = 2.1 [\mu m]$, which represents the properties of a highly porous aggregate. Both solutions fit the experimental data of the settling velocity equally well with $R^2 = 0.8$ collapsing onto a single curve. However, as can be seen in Figure \ref{fig:SK_all_fit}b, the resulting porosity differs over the entire range of the aggregate sizes $D$. Although the porosity values for both solutions are quite high ($0.97 < \epsilon < 1.0$), the difference in dry mass of the aggregates is significant. For the smallest floc analyzed by \cite{gillard2019} (i.e. for a floc with $D = 100 [\mu m]$) the dry mass of aggregates obtained by option 1 (lower porosity) is 6.5 times higher than than those obtained from option 2 (higher porosity). This ratio increases up to 23.9 for the largest flocs considered ($D=750 [\mu m]$). This comparison shows that the many degrees of freedom offered by the \textit{SK} model allow multiple parameter sets to provide equally good fits to the data for particles with very different properties in terms of mass and porosity, depending on the choice of parameters. The flexibility of the \textit{SK} model is a result of incorporating both the fractal theory relationships, which define the dependence between floc size and porosity, and the variable $\Omega$, which also depends on the size of the primary particles and aggregates.

It is now interesting to narrow down the degrees of freedom for the hydrodynamic properties by setting the drag coefficients constant as $C_1 = 20$ and $C_2 = 1$ to determine the ranges of $D_p$ and $n_f$ which result in the same $R^2$-value of 0.8. The resulting settling velocity for this test is shown in figure \ref{fig:SK_C1_fix}a. The model solution corresponding to the lowest porosity (Option A: $n_f = 2.19$, $D_p = 1 [\mu m]$) and satisfying the condition $R^2 = 0.8$ is shown as a semitransparent purple line. On the other hand, the model solution corresponding to the highest porosity is shown as an opaque purple line (Option B: $n_f = 1.55$, $D_p = 3.08 [\mu m]$). Comparing these two parameterizations of the \textit{SK} model, we see in figure \ref{fig:SK_all_fit}a that the plots differ only slightly in their slopes. The corresponding porosity for these two model results is given in figure \ref{fig:SK_C1_fix}b. These two lines provide the envelope of all possible model solutions with $R^2>0.8$ that are shown as the shaded purple area in figure \ref{fig:SK_C1_fix}. Even though the results of all possible solutions show less variation in porosity than the analysis in Figure \ref{fig:SK_all_fit}b, the difference still remains significant. For the Option A, the floc sizes $D=100 [\mu m]$ and $D=750 [\mu m]$ yield the porosities $\epsilon = 0.993$ and $\epsilon = 0.9997$, respectively. The latter, we evaluate as an unreasonably high value. On the other hand, for Option B, we obtain the respective porosities of $\epsilon = 0.962$ and  $\epsilon = 0.992$. This difference in porosity between model parameterisations results in the dry mass ratios of 3.5 for the smallest aggregate ($D= 100 [\mu m]$), and increases up to  13.6 for $D= 750 [\mu m]$. Therefore, it can be concluded that even if the drag coefficients $C_1$ and $C_2$ are treated as constants, there is still a range of $n_f$ and $D_p$ values that correspond to similar velocity plots but different aggregate porosities with potentially different hydrodynamic properties. 

From the analysis presented in figure \ref{fig:SK_C1_fix}, it is clear that the settling velocities can be tuned using different values of $n_f$ and $D_p$, which in turn changes the hydrodynamic properties $\epsilon$ and $\Omega$. To understand how each of these two hydrodynamic properties affects the model behaviour, we turn to the models \textit{FR} and \textit{PE} introduced in section \ref{sec:parameters}. This allows us to test the effects of $\epsilon$ and $\Omega$ individually, by fixing one of the components for the respective model and leaving only one degree of freedom for the \textit{FR} and \textit{PE} models.

\begin{figure*}[!b]
     \begin{center}
         \includegraphics[width=8cm]{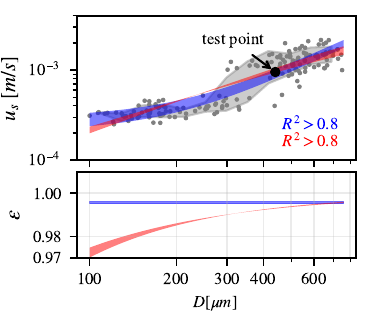}
         \put(-240,170){$a)$}
         \put(-240,80){$b)$}
     \end{center}
     
     \caption{
     Same as figure \ref{fig:SK_C1_fix}, but instead of the \textit{SK} model, \textit{FR} and \textit{PE} models are compared.  The shaded blue (\textcolor{trans_blue}{\rule[0.0mm]{0.5cm}{5pt}}) and red (\textcolor{trans_red}{\rule[0.0mm]{0.5cm}{5pt}}) areas in (a) and (b) represent all possible solutions of the \textit{PE} and \textit{FR} models that correlate with experimental data with $R^2 > 0.8$. The aggregate properties indicated by "test point" are used in the numerical simulations in section \ref{ch:num_sim}.
     }
\label{fig:FR_PE}
 \end{figure*}

By collecting all possible solutions of the \textit{FR} model with $\Omega=1$ and the variable $\epsilon$ as a function of $n_f$ and $D_p$ (equation Eq.\ref{eq:fractal_density_diff}) that yield $R^2 > 0.8$, we illustrate the model's settling velocity and corresponding porosity as the shaded red area in figure \ref{fig:FR_PE}a. Here the values of $C_1$ and $C_2$ are again set to 20 and 1.0 respectively, and the parameters $n_f$ and $D_p$ are in the ranges $2.0<n_f<2.2$, $1.0 [\mu m] < D_p < 2.97 [\mu m]$. The best fitting parameters are $n_f = 2.14$ and $D_p = 1.35 [\mu m]$ and the resulting $R^2 = 0.81$. The corresponding porosity range is shown in figure \ref{fig:FR_PE}b as a shaded red area. Comparing the width of this porosity range with that shown in figure \ref{fig:SK_C1_fix}b, it can be seen that the porosity varies less for the \textit{FR} model while producing similar $R^2$ values. It can also be seen that the resulting settling velocity and porosity of the \textit{FR} model shown in figure \ref{fig:FR_PE} are very close to the lower limit of the \textit{SK} model shown in figure \ref{fig:SK_C1_fix}b.

We now proceed to analyse the effect of $\Omega$ by collecting all possible solutions of the \textit{PE} model with $\epsilon=const$ and variable $\Omega$ as a function of $D_p$ and $\epsilon$. Note that $\epsilon$ becomes a fitting parameter for this model, which is no longer a function of floc size. As a result, the floc is no longer described by $n_f$. This assumption is motivated by the recent finding of \citet{spencer2021structure} that the internal structure of flocs may not be well represented by a fractal dimension. The shaded blue area in figure \ref{fig:FR_PE} shows all possible parameter combinations of the \textit{PE} model that give $R^2 > 0.8$. We follow the same approach as for the \textit{FR} model and set $C_1 = 20$ and $C_2 = 1.0$, which resulted in $0.9953<\epsilon<0.9961$, and $3.15 [\mu m] < D_p < 4.2 [\mu m]$. Similar to the porosity of the \textit{FR} model, figure \ref{fig:FR_PE}b also contains the shaded blue area corresponding to a high correlation ($R^2 > 0.8$). It can be seen that for the \textit{PE} model the porosity is higher than for the \textit{FR} model for all floc sizes analysed. Note that the porosity values for the \textit{FR} model approach those of the \textit{PE} model for larger floc sizes. The fact that the maximum porosity values are comparable between these two models is an advantage of the newly introduced \textit{PE} model, as it does not predict the extremely high porosities of the \textit{SK} model ($\epsilon = 0.9997$). Figure \ref{fig:FR_PE}a also shows that the settling velocity of the \textit{PE} model has a nonlinear trend, which reflects the sedimentation velocity plateau for small aggregates ($100 [\mu m] < D < 250 [\mu m]$). While the \textit{PE} model captures this trend, a similar plateau observed for larger aggregates ($500 [\mu m] < D < 750 [\mu m]$). The plateau observed for large aggregates was not captured by the models \textit{SK}, \textit{FC}, \textit{FR} and \textit{PE} and may be an interesting feature that deserves more attention in future studies. Note that such a plateau is not consistently reported for the other datasets analyzed in Appendix B.

The above analysis leads to the conclusion that the \textit{SK} model gives equally good fitting results for the settling velocity using different combinations of $D_p$ and $n_f$ resulting in different aggregate porosities. The variation in porosity is reflected in differences in dry mass, which can be up to an order of magnitude for the parameterisations that give the lightest and heaviest flocs. We note that while in this section we only present our analysis by comparison with the data from \citet{gillard2019}, in Appendix B we provide further comparisons of the \textit{FC}, \textit{FR} and \textit{PE} models, which all yield the same results.To rigorously define the models that predict the properties of the lightest and heaviest flocs, we propose to simplify the \textit{SK} model into two - \textit{FR} and \textit{PE}. Consequently, we use the best-fit results of the \textit{FR} and \textit{PE} models to obtain parameterisations that represent a heavy and a light floc, respectively, settling at the same rate in an unbounded flow. This is indicated by the test point in figure \ref{fig:FR_PE}. These two types of flocs will be investigated for more complicated flow configurations in the next sections.  

\subsection{Numerical simulations}
\label{ch:num_sim}

When comparing settling velocity models designed for individual aggregates that represent the dilute case, it can be seen that different assumptions for settling velocity models result in different properties of aggregates but similar settling behaviour. In this section, we use numerical simulations to examine the flow patterns around individual falling aggregates, whose properties correspond to \textit{PE} and \textit{FR} settling velocity models.  We also compare the settling dynamics of porous aggregates in more complex situations that bear much richer flow phenomena.

\begin{figure}[b!]
        \centering
        \includegraphics[width=11cm]{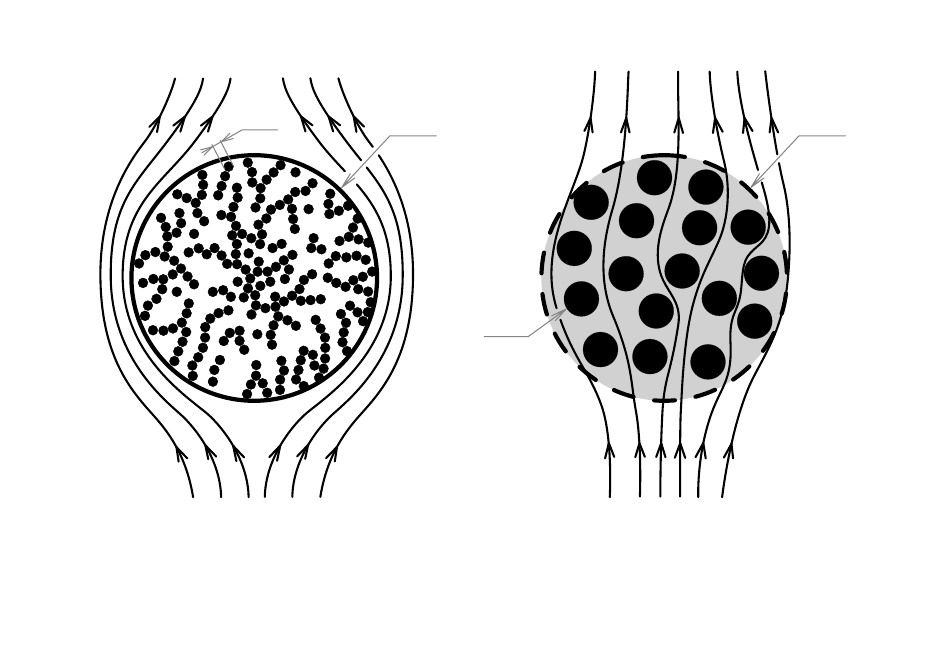}
        \put(-300,200){$a)$}
        \put(-150,200){$b)$}
        \put(-270,200){\textit{Non-permeable}}
        \put(-115,200){\textit{Permeable}}
        \put(-45,172){$D$}
        \put(-181,172){$D$}
        \put(-230,174){$D_p$}
        \put(-152,106){$D_p$}
        \caption{Illustration of the non-permeable \textit{a)} and permeable \textit{b)} aggregates that correspond to models \textit{FR} and \textit{PE}, respectively.} 
        \label{fig:flocs_scheme}
\end{figure}
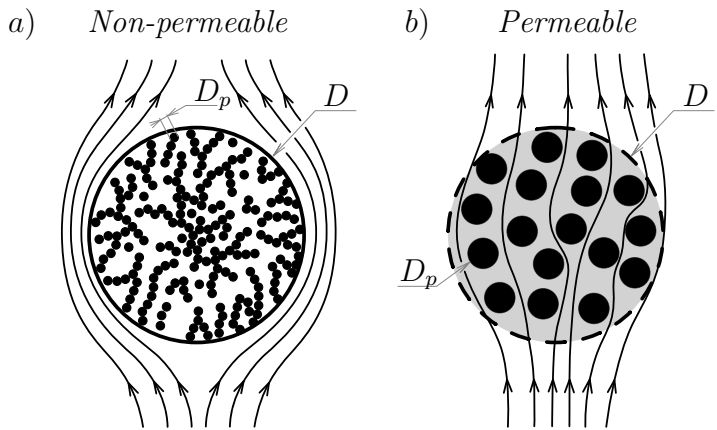

One prominent example of such a situation is the settling behavior of sediments in the dense regime that gives rise to hindered settling of suspensions with higher concentrations \citep{WINTERWERP20021339,richardson1954sedimentation}. For large enough concentrations, the particles experience a modified drag force due to the presence of wakes and return flows of neighboring particles or aggregates. As a result, the settling rate of sediments could be approximately estimated by the factor of $(1-\phi)$, where $\phi$ is the volumetric solid concentration  \citep{WINTERWERP20021339}. Another typical situation is the settling of porous aggregates in stratified fluids, where the change of fluid density, caused e.g. by a change in salinity or temperature, has to be considered. Such a scenario impacts the dynamics of highly porous aggregates such as marine snow \citep{prairie2013delayed,prairie2015delayed,panah2017simulations}. 

In order to compare the settling behavior in various conditions, we use DNS by resolving all scales of the fluid motion around the aggregates. We thereby capture numerically the geometry of aggregates that we simplify to spherical objects. In addition, our numerical approach considers the flow modification inside a homogeneous, porous material of aggregates, which is described in more detail in the next sections.

\begin{table}[t!]
    \centering
    \begin{tabular}{rl|cl|cl}

        \multicolumn{2}{c}{Parameters} & \multicolumn{2}{c}{\textit{non-permeable}} & \multicolumn{2}{c}{\textit{permeable}} \\
        \hline
        \hline
        $D$ & [$\mu m$] & 440 & \textit{fixed} & 440 & \textit{fixed}\\
        $\rho_s/\rho_f$ & [-]& 2.365 & \textit{fixed} & 2.365 & \textit{fixed}  \\
        \hline
        $D_p$ & [$\mu m$] & 1.55 & \textit{fitted} & 4.1 & \textit{fitted} \\
        $R_{agg}$ & [-]& $9.7\cdot 10^{-3}$ & $f(\epsilon,\rho_s)$ & $5.97\cdot 10^{-3}$ & \textit{fitted} \\
        $m_{dry}$ & [kg] & 7.93$\cdot{10}^{-10}$& $f(\epsilon,\rho_s)$ & 4.86$\cdot{10}^{-10}$ & $f(\epsilon,\rho_s)$\\ 
        $\epsilon$ & [-]& 0.99282 & $f(n_f,D_p)$ & 0.9956 & $f(R_{agg},R_s)$ \\
        $\beta$ & [-]& 100.0 & \textit{fixed} & 3.63 & $f(\epsilon,D,D_p)$ \\
        $\Omega$ & [-] & 0.99 & \textit{fixed} & 0.67 & $f(\epsilon,D,D_p)$ \\
       
    \end{tabular}
    \caption{Physical properties of the test particles taken for comparison. The values of aggregates properties are provided together with the description of the source for a certain quantity. The parameters labeled as \textit{fixed} are constants, \textit{fitted} parameters are adjusted to achieve the best-fit settling velocity, and the remaining parameters are derived. Parameter values labelled \textit{fitted} are taken from the analysis in section \ref{sec:fitting_with_experiments}.}
    \label{tab:test_particle_params}
\end{table}

To demonstrate the impact of porosity for the two aforementioned scenarios, we choose two sets of properties that correspond to the same settling velocity in undisturbed conditions as investigated in the previous sections. The size and settling velocity of these aggregates are indicated as a \textit{test point} in figure \ref{fig:FR_PE} and correspond to the properties of aggregates determined by the models  \textit{FR},  Eq.\ref{eq:sv_frac} and \textit{PE}, Eq.\ref{eq:sv_porous}. We label the aggregate that corresponds to the model \textit{FR}  as \textit{non-permeable}, because the low value of aggregate permeability is neglected for this settling velocity model. On the other hand, the aggregate that corresponds to model \textit{PE} is labeled as \textit{permeable}. Note that even though the permeability is neglected for the \textit{non-permeable} aggregate in the settling velocity model, the porous structure of the aggregate's material facilitates the process of diffusion. The aggregates are schematically illustrated in figure \ref{fig:flocs_scheme}, while their physical properties and the way we determine their values are summarized in table \ref{tab:test_particle_params}.

The main difference between these two aggregates is their dry mass and permeability. While the \textit{permeable} aggregate is lighter, it experiences less drag force while settling at the same rate due to the more permeable internal composition by comparison with the \textit{non-permeable} aggregate \sout{(\textit{cf}. table \ref{tab:test_particle_params})}. To mitigate discrepancies associated with numerical methods, we use a unified modeling approach to capture the behaviors of both aggregate types. We modeled the \textit{non-permeable} aggregate by applying very low permeability values $\kappa$, which renders the dimensionless permeability to be $\beta = 100$ and yields   $\Omega = 0.99$ according to Eq.\ref{eq:Omega}. This means that the aggregate experiences only about 1\% less drag force compared to a completely impermeable particle.

As a first step, we validate our code by simulating the settling of a single particle in unbounded flow. As expected, the simulation results reproduce the settling velocity of the experimental observation without any further tuning of the parameters (\textit{viz.} the settling velocity for the \textit{test point} in figure \ref{fig:FR_PE}a and the steady state velocity in figure \ref{fig:settling_por_vs_fractal}a). Then, we conducted simulations of two test cases for the two types of particles: (i) hindered settling and (ii) the settling in a density-stratified fluid. It is seen from table \ref{tab:test_particle_params} that the aggregates have the same diameter and density ratio $\rho_s/\rho_f$. However, due to the different porosities $\epsilon$, the dry mass of the aggregates $m_{dry}$ is considerably different with the \textit{non-permeable} particle being 63\% heavier. 

To present the results of numerical simulations in a non-dimensional way, we use the Stokes settling velocity Eq.\ref{eq:Stokes_settling_velocity} of an impermeable spherical aggregate $u_s^{St}$ as a reference velocity. The reference length scale is chosen to be equal to the aggregate diameter $D$. The reference time scale then becomes  $\tau = D/u_s^{St}$.  The Reynolds number is defined as $Re = \frac{u_s^{St}D}{\nu} = 0.538$ and kept constant for all numerical simulations.

\subsubsection{Settling behavior in fluids with constant density}
\paragraph{Governing equations and numerical method}
\leavevmode\newline
To perform numerical simulations with fluids of constant density, we compute the fluid motion by integrating the Navier-Stokes Equation (NSE) for incompressible fluids together with the continuity equation

\begin{eqnarray}
    \label{eq:NS_Darcy}
    \frac{\partial {\vec{u}}}{\partial t} + \vec{u} \cdot \nabla \vec{u} =& -\frac{1}{\rho_f} \nabla {p} + \nu \nabla^2 {\vec{u}} -\nu \kappa^{-1} \epsilon (\vec{u}-\vec{U}),\\
    \label{eq:continuity}
    \nabla \cdot \vec{u} = 0,
\end{eqnarray}

where $\vec{u}$, $t$, $p$, and $\vec{U}$ are the fluid velocity vector, time, pressure and the velocity vector of an aggregate, respectively. To account for the flow inside the porous medium of the aggregates, we add a penalization term to the Right-Hand Side (RHS) of the NSE that acts inside the aggregates only  \citep{panah2017simulations}. This penalization term is given as the last term on the RHS of  Eq.\ref{eq:NS_Darcy} and describes the resistance of aggregates on the velocity field in the presence of a porous medium. The current approach implies a smooth transition between the porous medium and the ambient fluid \citep{CHANDESRIS20073422}. 

To capture the dynamics of porous aggregates, we utilize the Newton-Euler equations for spherical particles:

\begin{eqnarray}
    \label{eq:part_dynamics}
    m \frac{d\vec{U}}{dt} =& \vec{F}_{h} + \vec{F}_{g} + \vec{F}_{c} \\
    I \frac{d\vec{\omega}}{dt} =& \vec{T}_{h} + \vec{T}_{c}
\end{eqnarray}

where $m = (\rho_s(1-\epsilon) + \rho_f \epsilon) V_{agg}$ is the total mass of an aggregate, $I$ and  $\omega$ are the moment of inertia and the angular velocity of an aggregate, respectively. Furthermore, $\vec{F}_{h}$, $\vec{F}_{g}$ and $\vec{F}_{c}$ are the hydrodynamic, gravity, and contact forces, while $\vec{T}_{h}$ and $\vec{T}_{c}$ are the torques due to hydrodynamic interactions as well as collisions and contacts. The hydrodynamic force is calculated via a volume integral of the last term of equation Eq.\ref{eq:NS_Darcy} over the volume of the aggregate $V_{agg}$, i.e. $\vec{F}_{h} = \int_{V_{agg}} \nu \kappa^{-1} \epsilon \rho_f (\vec{u}-\vec{U}) dV$. We use models for collisions and lubrication forces reported in literature \citep{biegert2017collision,vowinckel2019consolidation}. These models were validated for a wide range of Stokes numbers (up to $St = 152$) for particles represented by the Immersed Boundary Method (IBM). For the present simulation, the Stokes number based on the reference velocity of the \textit{non-permeable} aggregate $St = \rho_{agg} u_s^{St} D_p / 9 \rho_f \nu=0.052$ is relatively small. This suggests that aggregate interaction minimally influences hindered settling dynamics, allowing us to conclude that the collision and lubrication models derived for solid particles can serve as a suitable first approximation. 

\paragraph{Settling of a single aggregate}
\label{ch:simple_settling}
\leavevmode\newline

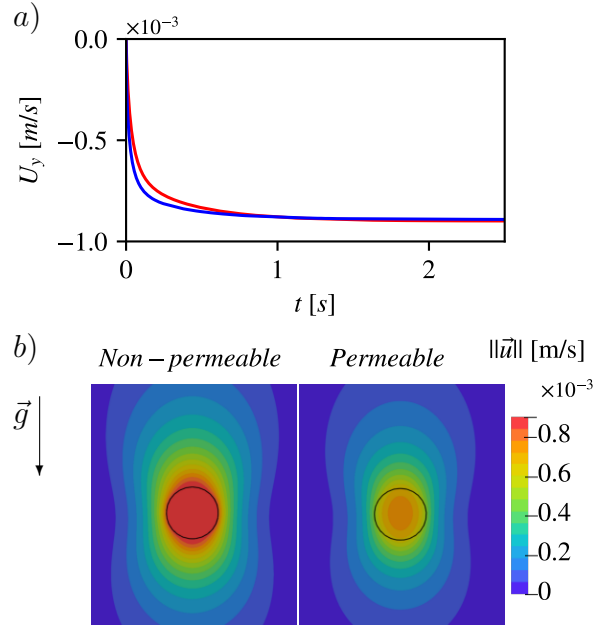
\begin{figure}[b!]
    \centering
    \includegraphics{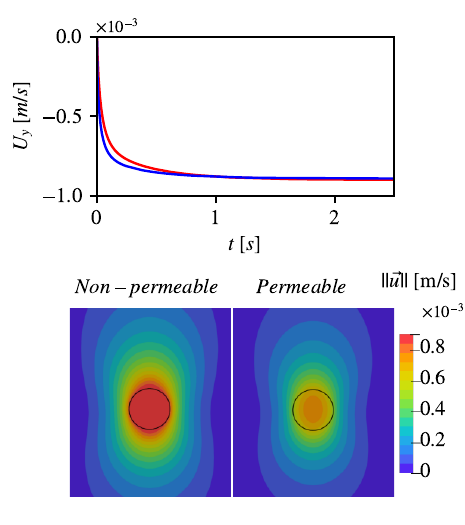}
    \put(-220,240){$a)$}
    \put(-220,115){$b)$}
    \put(-210,100){\vector(0,-1){30}}
    \put(-220,90){$\vec{g}$}
    \caption{The settling of aggregates: a) the settling velocity of the \textit{non-permeable} (\textcolor{red}{\rule[0.5mm]{0.5cm}{1pt}}) and \textit{permeable} (\textcolor{blue}{\rule[0.5mm]{0.5cm}{1pt}}) aggregates over time and b) 2D-contours of the flow fields cutting through the center of the aggregates at $t = 2 [s]$ settling under gravity ($\vec{g}$).
    } 
    \label{fig:settling_por_vs_fractal}
\end{figure}

To investigate the flow structure induced by single porous particles settling in a large container, we performed two simulations that correspond to the two different types of particles, \textit{viz.} the \textit{non-permeable} and \textit{permeable} particles introduced above with their properties summarized in table \ref{tab:test_particle_params}. The computational domain for this scenario is a cuboid domain of size $6.8D \times 20.45 D \times 6.8 D$. The origin of the Cartesian coordinate system is located at the left-back bottom corner of the computational domain. The aggregate is initialized at $P_0=(3.4D,18.75D,3.4D)$ with zero velocity in an otherwise quiescent fluid and the vector of gravity coincides with the negative $y$-direction. The resolution of the numerical mesh is $180\times540\times180$ cells, which yields 26.4 cells per aggregate diameter. The boundary conditions of the domain are free-slip walls for the vertical sides to minimize confining effects of the domain size and no-slip walls for the top and bottom plates. 

\begin{figure*}[b!]
    \centering
    \includegraphics[width=12cm]{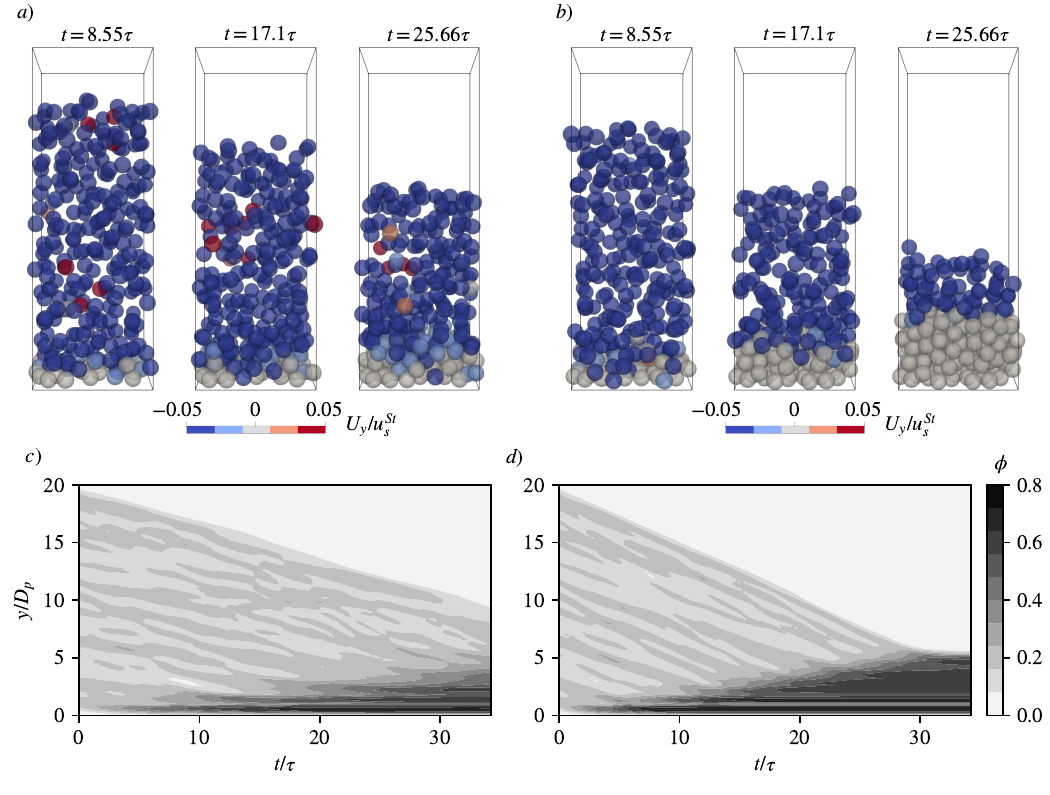}
    \caption{Visualization of instantaneous particle distributions for $a)$ \textit{non-permeable} and $b)$ \textit{permeable} aggregates and horizontally-averaged volume fraction $\phi$ over the height of a simulation domain as a function of time for $c)$ \textit{non-permeable}  and $d)$ \textit{permeable}  aggregates. Aggregates are visualized translucently on sub-figures $a)$ and $b)$ for better visibility. The narrow range for the coloring of the aggregates by their vertical velocity emphasizes the direction of particle movement rather than the velocity magnitude.} 
    \label{fig:hindered_snapshots}
\end{figure*}

Figure \ref{fig:settling_por_vs_fractal} shows the comparison between the \textit{non-permeable} and \textit{permeable} aggregate settling as individual objects. The vertical velocity is plotted over time in figure \ref{fig:settling_por_vs_fractal}$a$. Although slight differences are observed in the initial acceleration phase, the settling velocity converges to the same value for $t>1$~[$s$] for both types of particles. At this stage, however, the resulting velocity fields in the immediate vicinity of these particles are very different as illustrated in figure \ref{fig:settling_por_vs_fractal}$b$. This subfigure shows the instantaneous vertical velocity field at the well-developed stage for the settling ($t= 2$ [$s$]). For the  \textit{non-permeable} particle, it is seen that the vertical velocity of the fluid is constant within the particle volume and coincides with the vertical velocity of the particle. This was expected because the \textit{non-permeable} particle acts as a solid object that moves with constant velocity. For the \textit{permeable} particle, on the other hand, a lower value of fluid velocity is observed inside the aggregate that indicates the flow through the permeable object. Consequently, the wake behind the \textit{permeable} aggregate is smaller by comparison to its \textit{non-permeable} counterpart. Despite those differences in the local hydrodynamic flow patterns in the vicinity of the aggregates, the settling velocities for those aggregates are similar as predicted by the models \textit{FR} and \textit{PE} in figure \ref{fig:FR_PE}.

\paragraph{Hindered settling}
\label{ch:hindered_settling}
\leavevmode\newline

In contrast to the previous example, the settling of many aggregates is more complex as it depends on the flow patterns of neighboring settling aggregates. To study the difference in settling behavior for such a scenario, we performed numerical simulations in a computational domain of size $7D\times20D\times7D$, where the gravitational vector coincides with the negative $y$-direction. The  numerical grid comprises $182\times540\times182$ grid cells, which yields a resolution of 25.7 cells per particle diameter. The origin of the coordinate system is set to the left-back bottom corner of the numerical domain. We performed two simulations for the two types of particles \textit{non-permeable} \, and \textit{permeable} \, the properties of which are defined in table \ref{tab:test_particle_params}. The simulation scenario consists of 290 aggregates of equal size $D$ that are initially randomly distributed over the entire domain. This corresponds to a volume fraction of $\phi = V_{ags}/V_{dmn} = 0.155$, where $V_{ags}$ and $V_{dmn}$ are the volume occupied by the aggregates and the total volume of the simulation domain, respectively. The initial distribution of aggregates for both simulations is identical. The aggregates are initialized with zero velocity in a quiescent fluid and subsequently settle under gravity toward the bottom of the domain. Boundary conditions of the top and bottom of the domain are set as no-slip walls, whereas periodic boundary conditions are applied in the horizontal directions.

The time evolution of the horizontally averaged volume fraction as a function of height is presented in figure \ref{fig:hindered_snapshots}c and \ref{fig:hindered_snapshots}d. It is seen that lighter \textit{permeable} aggregates settle faster by comparison with heavier \textit{non-permeable} aggregates, which seems counter-intuitive. The explanation for this behavior can be found in the local hydrodynamic effects of  the \textit{permeable} aggregates. As can already be seen in figure \ref{fig:settling_por_vs_fractal}b, the two types of particles induce very different types of wake flows during their settling. Consequently, the counterflows in hindered settling are very different as well. While the non-porous \textit{non-permeable} particles force the fluid underneath to move upwards as they replace fluid volume \citep{vowinckel2019consolidation}, the counter flows of porous aggregates partially take place inside the particles, rendering the local upward currents less energetic. Hence, particles are exposed to weaker counterflows which accelerates the settling process. The effect of counterflows is demonstrated in figures \ref{fig:hindered_snapshots}$a$ and $c$, where aggregates are colored according to their vertical velocity. For the simulations of \textit{non-permeable} aggregates, it is seen that some of the aggregates move upwards due to the influence of counterflows (figure \ref{fig:hindered_snapshots}$a$). However, this effect is completely suppressed for the hindered settling of \textit{permeable} aggregates investigated here (figure \ref{fig:hindered_snapshots}$b$.)

\subsubsection{Settling behavior in fluids with  density stratification}
\label{ch:stratified_settling}
\paragraph{Governing equations and numerical method}
\leavevmode\newline
We now consider the settling of aggregates in a stably stratified environment to elucidate the effect of the permeability of aggregates on the transport of scalar quantities such as temperature or salinity. We account for this effect by introducing a non-dimensional concentration field $c$ ranging from zero to one that corresponds to an excess density, such that the density of the fluid is  $\rho_f = \rho_b \cdot c + \rho_u \cdot(1-c)$, where $\rho_b$ and $\rho_u$ are the density values of the bottom and upper layers of the domain, respectively. Due to the small difference between the lighter and denser fluid regions ($\rho_b/\rho_u < 1.05$), an additional buoyancy term $(\rho_f / \rho_u - 1)\cdot \vec{g} $ is introduced for the right-hand side (RHS) of the NSE that does not alter the fluid density in terms of fluid inertia \citep{vowinckel2021incorporating}. For the simulations with a density-stratified environment, we employ the moving frame of reference (MFR) that moves with the velocity of the settling aggregate. To this end, we introduce an additional term to the RHS of the NSE that accounts for the  momentum change in the domain due to the acceleration of the aggregate. This yields

\begin{equation}
    \label{eq:NS_Darcy_conc}
    \begin{split}
    \frac{\partial {\vec{u}}}{\partial t} + \vec{u} \cdot \nabla \vec{u} = -\frac{1}{\rho_u} \nabla {p} + \nu \nabla^2 {\vec{u}} -\nu \kappa^{-1} \epsilon (\vec{u}-\vec{U}) + \\
    + \frac{\left( \rho_f - \rho_u \right)}{\rho_u} \cdot \vec{g} + \frac{\left( \rho_f - \rho_u \right)}{\rho_u} \frac{d \vec{U}}{dt}.
    \end{split}
\end{equation}

To simulate concentration transport, we additionally solve a convection-diffusion transport equation:

\begin{equation}
    \label{eq:scalar_transport}
    \frac{\partial c}{\partial t} + \vec{u} \: \nabla {c} = {D_c} \nabla^2 {c},
\end{equation}
where $D_c$ is the molecular diffusion coefficient. Considering the high porosity of aggregates, i.e. $\epsilon \rightarrow 1$, and the relationship between diffusion within the porous media of aggregate $D_{eff}$ and the aggregate's porosity $\epsilon$ as expressed by $D_{eff} = D_c \epsilon^2$ \citep{manheim1974diffusimetry, yong2012environmental},  the approximation  $D_{eff} = D_c$ appears to be adequate. We solve  Eq.\ref{eq:NS_Darcy_conc} and Eq.\ref{eq:scalar_transport} together with  Eq.\ref{eq:part_dynamics}  for particle motion to model the dynamics of permeable aggregates settling in a  density-stratified fluid using the same schemes applied by \citet{biegert2017collision} and \citet{kollner2020gravity}. In addition to the Reynolds number, the presence of a scalar field yields additional non-dimensional parameters. Namely, the ratio between inertia and diffusivity is quantified by the Peclet number $Pe = \frac{u_s^{St}D}{2D_c}$, the density difference of the fluid stratification
is represented via $\eta = (\rho_b-\rho_u)/\rho_u$, and the density difference between the fluid and the particle is described by $\xi = \frac{\rho_b-\rho_u}{(\rho_s-\rho_u)(1-\epsilon)}$. The numerical procedure has been validated with high accuracy against experimental data (cf. figure A10 in Appendix A).

\paragraph{Settling of a single aggregate}
\leavevmode\newline

In contrast to the simplified scenario of a single settling aggregate, where the heavier \textit{non-permeable} aggregate develops the same settling speed as the \textit{permeable} one, the settling process in a  stably stratified fluid bears additional mechanisms that affect the settling dynamics. When an aggregate settles through the layer of lighter fluid towards a dense fluid layer, it transports lighter fluid  from the upper layer that is trapped inside the porous medium and increases the buoyancy of the aggregate. Depending on the porosity of an aggregate and the density difference of the fluid layers, the buoyancy can decrease drastically for a certain amount of time, thereby reducing its settling speed, which is also referred to as Delayed Settling Time (DST). The DST reflects the time scale for lighter fluid to be replaced by denser fluid inside the aggregate.
Note that in the limiting case of $\beta \to \infty$, there would not be any mass exchange between the aggregate and the ambient fluid, and the aggregate would stay at a density horizon that prescribes neutrally buoyant conditions. The  delay in settling within the density-stratified conditions was consistently studied with the application of marine snow dynamics as it plays an important role in governing marine carbon fluxes  \citep{kindler2010diffusion,prairie2015delayed,panah2017simulations,ahmerkamp2022settling}.

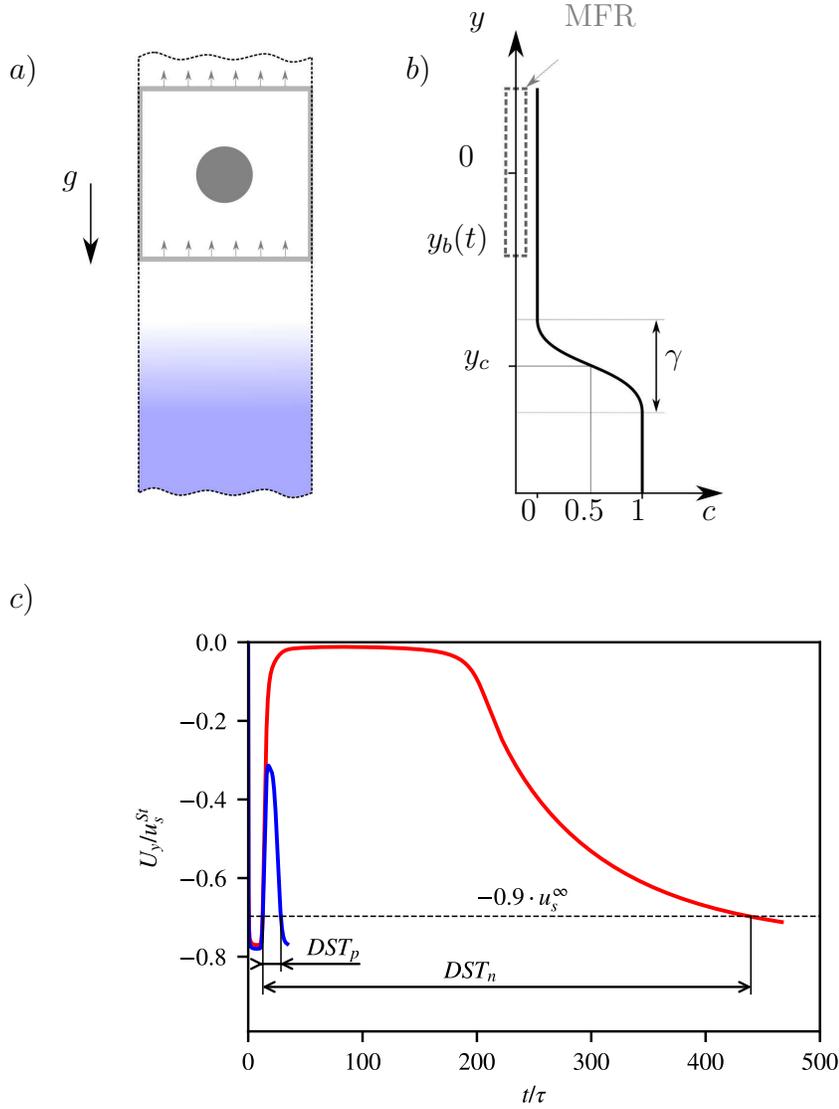
\begin{figure}[t!]
    \centering
    \includegraphics[width=8.5cm]{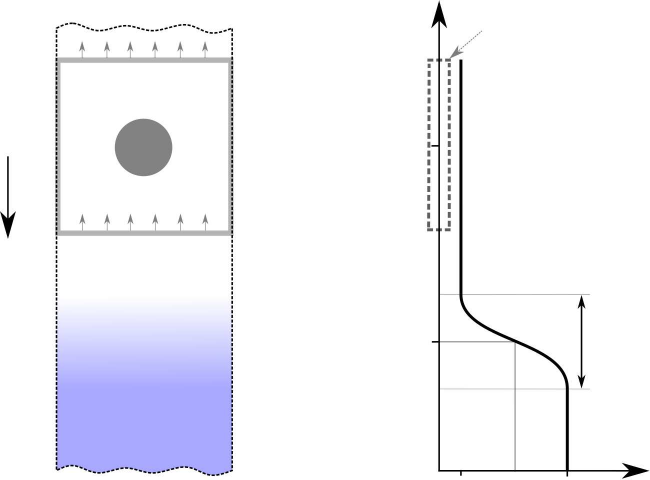}
    \put(-250,120){$g$}
    \put(-96,180){$y$}
    \put(-112,96){$y_b(t)$}
    \put(-8,-7){$c$}
    \put(-76.5,-7){$0$}
    \put(-60,-7){$0.5$}
    \put(-35,-7){$1$}
    \put(-22,52){$\gamma$}
    \put(-98,52){$y_c$}
    \put(-100,127){$0$}
    \put(-270,160){$a)$}
    \put(-120,160){$b)$}
    \put(-270,-40){$c)$}
    \put(-60,180){\textcolor{gray}{MFR}}
    \newline
    \vspace{0pt}
    \includegraphics[width=10.cm]{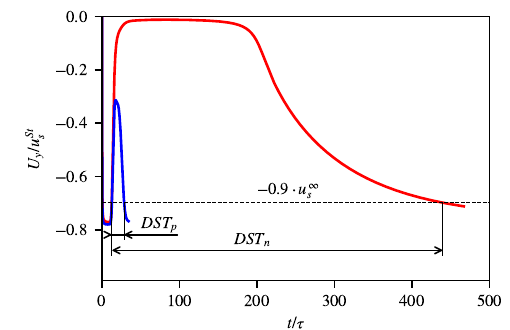}
    \caption{ Simulation setup and results of an aggregate settling in a density stratified fluid: $a)$ schematic of the computational domain in a moving frame of reference (MFR), where colors represent the concentration field, $b)$ the corresponding non-dimensional concentration profile over depth and $c)$ simulation results of the settling velocity of a \textit{non-permeable} (\textcolor{red}{\rule[0.5mm]{0.5cm}{1pt}}) and a \textit{permeable} (\textcolor{blue}{\rule[0.5mm]{0.5cm}{1pt}}) aggregate. Here, $U_y$ denotes the settling speed of the aggregate and $u_s^{\infty}$ corresponds to the magnitude of the steady settling velocity in the upper, lighter fluid layer. $DST$ is the Delayed Settling Time and the subscripts indicate DSTs for the \textit{non-permeable} and the \textit{permeable} aggregate, respectively.}
    \label{fig:delayed_settling_setup}
\end{figure}


To simulate the DST of the different aggregates, we define a cuboidal computational domain that includes a particle at the center coordinate of the domain at time $t=0$. The computational domain is represented via the MFR (solid grey lines in figure \ref{fig:delayed_settling_setup}$a$). The aggregate remains at its original position of the moving frame which is translated with the settling speed of the aggregate. The concentration value on the bottom of the domain is prescribed as a function of the bottom vertical position $y_b$ as follows:

\begin{equation}
    \label{eq:conc_init}
    c(y_b) = \frac{1-erf\left(\frac{4(y_{b}-y_c)}{\gamma} \right)}{2}
\end{equation}

where $y_c$ is the center coordinate of the concentration transition layer and $\gamma$ is the width of the density transition layer (figure \ref{fig:delayed_settling_setup}$b$). It is important to note that $y_b$ is a function of time as the MFR moves downwards.
The upper boundary conditions for velocity and concentration are similar and represent a convective outflow $D \vec{u}/Dt = 0$ and $Dc/Dt = 0$, respectively. The concentration field and the fluid velocity in the domain are initialized with zero values. The boundary conditions for the velocity on the side walls are set as free-slip, while for the bottom side of the domain, the velocity is set to be matching with the settling velocity of the falling aggregate.

To demonstrate the influence of particle permeability on the settling behavior and the DST in particular, we performed two simulations similar to the test case of single aggregates settling in a large container. We, again, used the properties of the \textit{non-permeable} and \textit{permeable} aggregates from table \ref{tab:test_particle_params}. The simulation domain represents a cuboid of size $6D\times6D\times6D$ with a particle initialized at the center of the domain.  The mesh resolution of the domain is $120\times120\times120$ cells or 20 cells per particle diameter. The origin of the Cartesian coordinate system is initialized in the middle of the aggregate position.  The numerical simulations are parameterized with $Pe=1746.0$, $\eta=0.0108$, $\gamma= 5.8D$, $\xi=1.8$ and $y_c = -9.25D$ that correspond to experimental conditions of marine snow settling in a density-stratified environment \citep{camassa2013retention}.

Figure \ref{fig:delayed_settling_setup}$c$ shows the settling speeds of the \textit{non-permeable} and the \textit{permeable} aggregate normalized with its respective Stokes settling velocity. As expected from the previous analyses, the settling speeds are identical for the two types of particles in the acceleration and the steady settling phase ($t < 7\tau$) as aggregates pass through the lighter fluid. When aggregates enter the region of the density transition $\gamma$, they start to decelerate at approximately the same rate. However, the \textit{permeable} aggregate recovers its original settling velocity faster than the \textit{non-permeable} aggregate, \textit{viz.} undergoes a shorter DST, as the lighter fluid in the aggregate is replaced much quicker by the surrounding denser fluid. The faster replacement is caused by the convective part of scalar transport inside the \textit{permeable} aggregate. In contrast, the settling speed of the impermeable \textit{non-permeable} aggregate decreases drastically and reaches almost zero. As time progresses, the heavier fluid diffuses into the aggregate by molecular diffusion, returning it to the original settling velocity. As a result, the ratio of the DSTs for \textit{non-permeable} ($DST_n$) and \textit{permeable} ($DST_p$) aggregates is $DST_n/DST_p = 26.97$. Even though this ratio depends also on the molecular diffusivity of a scalar field, the huge difference in DSTs highlights the importance of determining the permeable properties of aggregates in order to predict the sediment flux of porous and permeable aggregates.
Note, that while in Appendix A we showed that the reduced diffusion coefficient inside the aggregate showed a better comparison with the experimental results, within the present simulation we kept the diffusion coefficient inside aggregates the same as for the free fluid phase. The decreased diffusivity inside the aggregate would result in even larger difference between the $DST_n$ and $DST_p$.

\section{Conclusions and outlook} \label{sec:conclusions}
We have reviewed several commonly used settling velocity models and derived two simplified versions thereof to compare to various experimental measurements. The first simplified model takes the density variation over the aggregate size into account by using the assumptions of the fractal geometry of the aggregates. The second simplified model neglects the density change but considers the permeability of an aggregate as the main factor affecting the settling velocity. Our analysis showed that both models can capture experimental results of single settling aggregates with the same level of accuracy. However, the properties of the aggregates that correspond to these models can be drastically different, which can have strong implications for the settling behavior in more complex situations.
For example, we demonstrate that the aggregate dry mass predicted by the model that disregards permeability can be up to an order of magnitude higher than that of aggregates whose properties were determined using the model that incorporates permeability.

The difference in aggregate properties is demonstrated for the two prototype examples of settling in stratified environments and hindered settling, where permeability is key to predicting the settling behavior of porous aggregates. We showed that hindered settling behavior is sensitive to the permeability of the aggregates. Hindered settling of permeable aggregates reduces the energy of upstream counter-flows that are induced by the displacement of fluid by settling particles. This yields the counter-intuitive behavior that lighter, permeable aggregates settle faster than denser, impermeable structures. Furthermore, for settling in stratified environments, the displacement of the lighter fluid within the porous aggregate by the denser phase is governed by diffusion and convection transport. While convection transport is minimized for impermeable aggregates, we show that diffusion is the main mechanism to alter the Delayed Settling Time and, hence, the residence time of the porous aggregate in a stratified water column. It is worth noting that the simulations of the density-stratified environment were carried out with the same diffusion coefficients between the dense and light fluids inside and outside the aggregate. However, in Appendix A we show that a reduced diffusivity inside the aggregate results in a better agreement with experimental observations. Therefore, we acknowledge that the possible variation of the diffusion coefficient inside the porous aggregates could potentially improve the accuracy of this numerical approach even further and hence deserves more investigations in the future.

The present study therefore clearly demonstrates the importance of determining the exact properties of natural aggregates. We show that aggregates properties cannot be estimated if only the settling velocity of flocs with the dependency of their size is measured experimentally. The analysis of settling dynamics in a density-stratified environment or hindered settling could be used in order to estimate the permeability and porosity of aggregates. Future work should therefore focus on a more comprehensive description of the porous aggregates in terms of their permeability and shape. These efforts will help to further improve, among other things,  macroscopic sediment transport models in which the settling velocity of suspended sediment is a key component.


\appendix

\section{Validation of simulations of particles settling in a density-stratified environment}
\label{ap:validation}

\begin{figure}[b!]
    \centering
    \includegraphics[width=12cm]{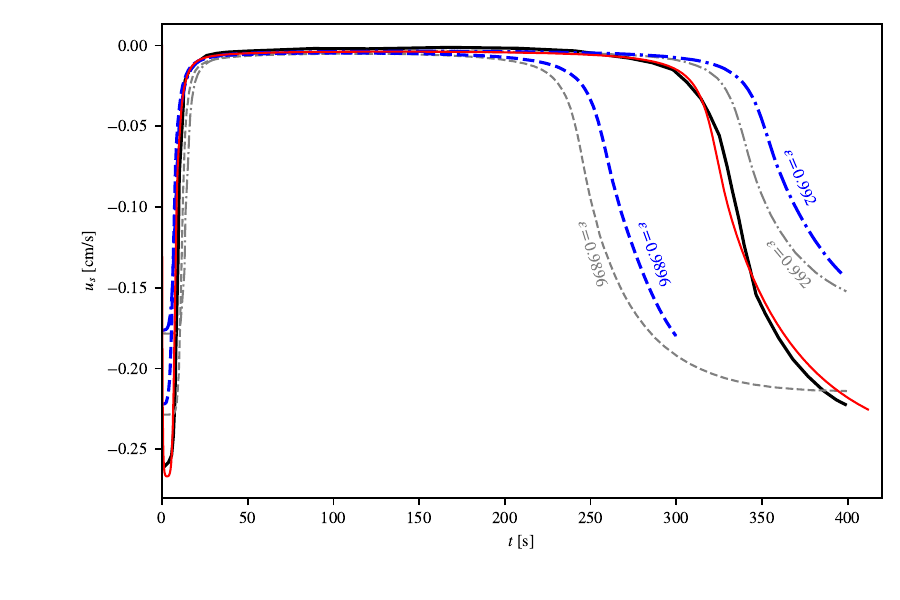}
    \caption{Settling velocity of aggregates over time. Curve \textcolor{black}{\rule[0.5mm]{0.5cm}{1pt}} corresponds to the experimental measurements of \cite{camassa2013retention}. Curves \textcolor{gray}{\rule[0.5mm]{0.5cm}{1pt}} correspond to the numerical simulations from \cite{panah2017simulations} for different values of porosity of the aggregate ($\epsilon = 0.9896$ and $\epsilon = 0.992$). Curve \textcolor{blue}{\rule[0.5mm]{0.5cm}{1pt}} corresponds to the present simulations performed for the same porosity values as were used in \cite{panah2017simulations}. Curve \textcolor{red}{\rule[0.5mm]{0.5cm}{1pt}} correspond to the porosity of the aggregate $\epsilon = 0.9888$ and modified diffusion coefficient inside the aggregate $D_c^f = 0.741 D_c$}
    \label{fig:delayed_settling_validation}
\end{figure}

In order to validate our numerical model, we performed several simulations of a settling aggregate with properties adopted from the experimental study of  \cite{camassa2013retention}. We compared the aggregate settling velocity over time with experimental \citep{camassa2013retention} and numerical \citep{panah2017simulations} results of the same settling scenario. The domain size is $4D\times75D\times4D$ with a resolution of $120\times2250\times120$, which corresponds to 30 cells per aggregate diameter. The simulation parameters are the following: $Re = 2.2$,
$Pe = 1746$, $\eta = 0.023$, $\gamma = 23.2 D_p$, $\xi = 5.5$ and $y_c = -24 D_p$. For the present test case we did not use the moving frame of reference. In addition, similar to numerical simulations reported in \cite{panah2017simulations}, we suppress the convective part of the concentration transport inside  porous aggregates. This implies using the following equation for the concentration field:

\begin{equation}
    \label{eq:scalar_transport_noconv}
    \frac{\partial c}{\partial t} + (1-M) \vec{u} \: \nabla {c} = {D_c} \nabla^2 {c},
\end{equation}
where $M$ is an indicator function that is set to unity inside an aggregate and equals zero otherwise. The boundary conditions for the velocity are free-slip on the boundary surfaces that are parallel to the gravity vector and no-slip on the boundary surfaces that are perpendicular to the gravity vector. 

The velocity and pressure are initialized with zero value, while the concentration $c$ is initialized as the function of the vertical $y$ component:
\begin{equation}
    \label{eq:conc_init_2}
    c(y) = \frac{1-erf\left(\frac{4(y-y_c)}{\gamma} \right)}{2} .
\end{equation}
The boundary conditions for the concentration field and pressure are zero-gradient for all sides of the computational domain.

Figure \ref{fig:delayed_settling_validation} shows the comparison of the vertical velocity of an aggregate as a function of time for the current simulations and previous experimental and numerical data \citep{camassa2013retention,panah2017simulations}. \cite{panah2017simulations} provided two simulation results that correspond to different values of porosity. The value of porosity $\epsilon=0.992$ was adopted from the original experimental work of \cite{camassa2013retention}. As far as the Reynolds number $Re=2.2$ does not correspond to  Stokes flow, \citet{panah2017simulations}  calculated a different value for porosity $\epsilon=0.9896$ that corresponds to the empirical drag coefficient which is valid for a wider range of Reynolds numbers. The authors showed that the experimental value of DST lies in the range between these two estimated values of porosity. It is also seen from their data that neither of these porosity values captures the settling velocity that is reached before entering the density transition region ($t<25$s). Our simulation results for the same aggregate porosities yield slightly different settling velocities and DSTs. We explain this deviation from numerical simulations of \citet{panah2017simulations} by the fact that our approach is a 3-dimensional simulation with a slightly coarser resolution than the 2-dimensional simulations of \cite{panah2017simulations}. In our simulation, we chose the porosity of the aggregate in order to capture the settling velocity of the experiments $\epsilon = 0.9888$. However, by using this value of porosity we were unable to capture the DST of experiments. Hence, we performed an additional simulation with an enhanced mathematical description of the diffusion process of the scalar field. Inside the aggregate, we prescribe a  modified value of diffusion coefficient of $D_c^a = 0.741 D_c$, which yields a good match of our simulation results and the experimentally measured trajectory of a settling aggregate. This allows us to reproduce the DST  and the settling velocity reported by \cite{camassa2013retention} with high fidelity.
\newpage
\newlength{\figwidth}
\setlength{\figwidth}{9cm}
\section{Analysis of additional measured data of settling velocities from literature}
\label{ap:more_exp_data}
\begin{figure}[!ht]
    \centering
    \includegraphics[width = \figwidth]{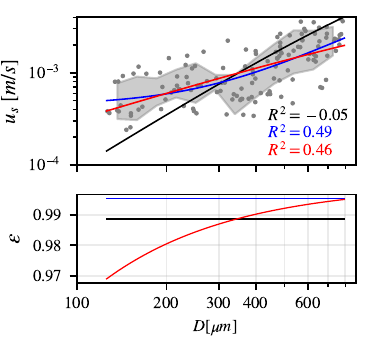}
    \put(-270,220){$a)$}
    \put(-270,90){$b)$}
    \caption{Settling velocity \textit{a)} and porosity \textit{b)} as a function of aggregate size of the of the \textit{FC} (\textcolor{black}{\rule[0.5mm]{0.5cm}{1pt}}), \textit{PE} (\textcolor{blue}{\rule[0.5mm]{0.5cm}{1pt}}), and \textit{FR} (\textcolor{red}{\rule[0.5mm]{0.5cm}{1pt}}) models compared with the experimental data set of \citet{gillard2019}. The experimental data (\textcolor{gray}{$\bullet$}) correspond to the sediment concentration of 105 [$mg/l$] and zero shear rate during flocculation. The grey area (\textcolor{mygray}{\rule[0.0mm]{0.5cm}{5pt}}) in (a) represents the normal distribution of the experimental data with a cut-off of the standard deviation $\sigma$.}
    \label{fig:ben_0_105}
\end{figure}

\begin{figure}[!ht]
    \centering
    \includegraphics[width = \figwidth]{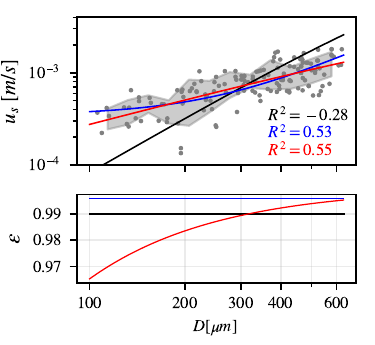}
    \put(-270,220){$a)$}
    \put(-270,90){$b)$}
    \caption{Settling velocity \textit{a)} and porosity \textit{b)} as a function of aggregate size of the of the \textit{FC} (\textcolor{black}{\rule[0.5mm]{0.5cm}{1pt}}), \textit{PE} (\textcolor{blue}{\rule[0.5mm]{0.5cm}{1pt}}), and \textit{FR} (\textcolor{red}{\rule[0.5mm]{0.5cm}{1pt}}) models compared with the experimental data set of \citet{gillard2019}. The experimental data (\textcolor{gray}{$\bullet$}) correspond to the sediment concentration of 175 [$mg/l$] and zero shear rate during flocculation. The grey area (\textcolor{mygray}{\rule[0.0mm]{0.5cm}{5pt}}) in (a) represents the normal distribution of the experimental data with a cut-off of the standard deviation $\sigma$.}
    \label{fig:ben_0_175}
\end{figure}

\begin{figure}[!ht]
    \centering
    \includegraphics[width = \figwidth]{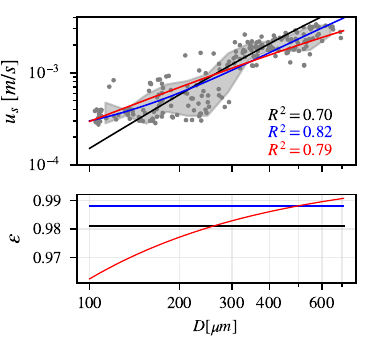}
    \put(-270,220){$a)$}
    \put(-270,90){$b)$}
    \caption{Settling velocity \textit{a)} and porosity \textit{b)} as a function of aggregate size of the of the \textit{FC} (\textcolor{black}{\rule[0.5mm]{0.5cm}{1pt}}), \textit{PE} (\textcolor{blue}{\rule[0.5mm]{0.5cm}{1pt}}), and \textit{FR} (\textcolor{red}{\rule[0.5mm]{0.5cm}{1pt}}) models compared with the experimental data set of \citet{gillard2019}. The experimental data (\textcolor{gray}{$\bullet$}) correspond to the sediment concentration of 105 [$mg/l$] and $2.4 [s^-1]$ shear rate during flocculation. The grey area (\textcolor{mygray}{\rule[0.0mm]{0.5cm}{5pt}}) in (a) represents the normal distribution of the experimental data with a cut-off of the standard deviation $\sigma$.}
    
    \label{fig:ben_2.4_105}
\end{figure}

\begin{figure}[!ht]
    \centering
    \includegraphics[width = \figwidth]{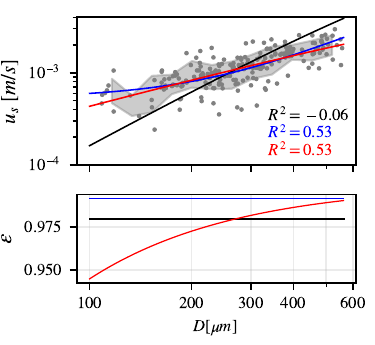}
    \put(-270,220){$a)$}
    \put(-270,90){$b)$}
    \caption{Settling velocity \textit{a)} and porosity \textit{b)} as a function of aggregate size of the of the \textit{FC} (\textcolor{black}{\rule[0.5mm]{0.5cm}{1pt}}), \textit{PE} (\textcolor{blue}{\rule[0.5mm]{0.5cm}{1pt}}), and \textit{FR} (\textcolor{red}{\rule[0.5mm]{0.5cm}{1pt}}) models compared with the experimental data set of \citet{gillard2019}. The experimental data (\textcolor{gray}{$\bullet$}) correspond to the sediment concentration of 175 [$mg/l$] and $2.4 [s^-1]$ shear rate during flocculation. The grey area (\textcolor{mygray}{\rule[0.0mm]{0.5cm}{5pt}}) in (a) represents the normal distribution of the experimental data with a cut-off of the standard deviation $\sigma$.}
    \label{fig:ben_2.4_175}
\end{figure}

\begin{figure}[!ht]
    \centering
    \includegraphics[width = \figwidth]{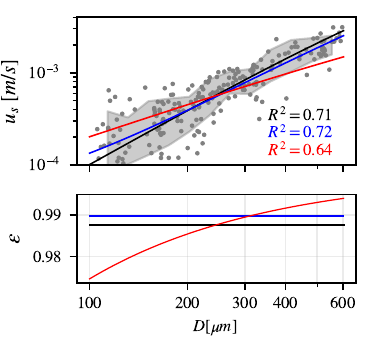}
    \put(-270,220){$a)$}
    \put(-270,90){$b)$}
    \caption{Settling velocity \textit{a)} and porosity \textit{b)} as a function of aggregate size of the of the \textit{FC} (\textcolor{black}{\rule[0.5mm]{0.5cm}{1pt}}), \textit{PE} (\textcolor{blue}{\rule[0.5mm]{0.5cm}{1pt}}), and \textit{FR} (\textcolor{red}{\rule[0.5mm]{0.5cm}{1pt}}) models compared with the experimental data set of \citet{gillard2019}. The experimental data (\textcolor{gray}{$\bullet$}) correspond to the sediment concentration of 500 [$mg/l$] and $2.4 [s^-1]$ shear rate during flocculation. The grey area (\textcolor{mygray}{\rule[0.0mm]{0.5cm}{5pt}}) in (a) represents the normal distribution of the experimental data with a cut-off of the standard deviation $\sigma$.}
    \label{fig:ben_2.4_500}
\end{figure}

To supplement section \ref{sec:fitting_with_experiments}, figures \ref{fig:ben_0_105}--\ref{fig:ben_2.4_500} provide  additional comparisons of experimental measurements reported in \cite{gillard2019} and various settling velocity models that correlate to the respective experimental data. Tables \ref{tab:ben_0_105}--\ref{tab:ben_2.4_500} report the best-fit parameters that correspond to each settling velocity model within each corresponding data set. The density of the ambient fluid, the density of solid aggregate material, and the fluid viscosity are constants and are reported in the main body of the current paper. The ranges for the fitting parameters are also reported in Table 1 of the main body of the current paper. The grey area represents the normal distribution of the corresponding experimental data with a cut-off of the standard deviation $\sigma$. The same formatting also holds for all figures in this section of the appendix.

Figures \ref{fig:man_kaolinite}--\ref{fig:man_kaol_bent} provide the measurement data for the settling of suspensions comprising Kaolinite, Bentonite, and a mixture of both materials, respectively. Within this comparison, we assume a constant density of the solid material $\rho_s = 2650$ [kg/m$^3$], $\rho_f = 1028$ [kg/m$^3$], $\nu$ = 1.05 $\times$ 10$^{-6}$ [m$^2$ s$^{-1}$]. We also change the maximal limit of the primary particle size $10^{-6}<D_p<5\times 10^{-5}$. Tables \ref{tab:man_kaolinite}--\ref{fig:man_kaol_bent} report the best-fit parameters that correspond to each settling velocity model within a corresponding data set.

\begin{figure}[!ht]
    \centering
    \includegraphics[width = \figwidth]{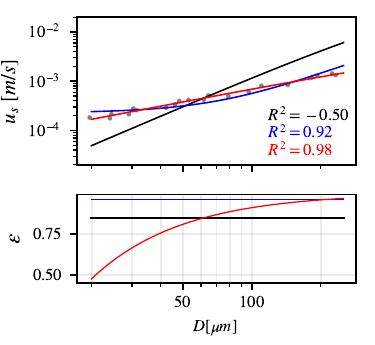}
    \put(-270,220){$a)$}
    \put(-270,90){$b)$}
    \caption{Settling velocity \textit{a)} and porosity \textit{b)} as a function of aggregate size of the of the \textit{FC} (\textcolor{black}{\rule[0.5mm]{0.5cm}{1pt}}), \textit{PE} (\textcolor{blue}{\rule[0.5mm]{0.5cm}{1pt}}), and \textit{FR} (\textcolor{red}{\rule[0.5mm]{0.5cm}{1pt}}) models compared with the experimental data set of \citet{gillard2019}. The experimental data (\textcolor{gray}{$\bullet$}) correspond to the study of \citet{gibbs1985estuarine}. The grey area (\textcolor{mygray}{\rule[0.0mm]{0.5cm}{5pt}}) in (a) represents the normal distribution of the experimental data with a cut-off of the standard deviation $\sigma$.}
    \label{fig:gibbs}
\end{figure}

\begin{figure}[!ht]
    \centering
    \includegraphics[width = \figwidth]{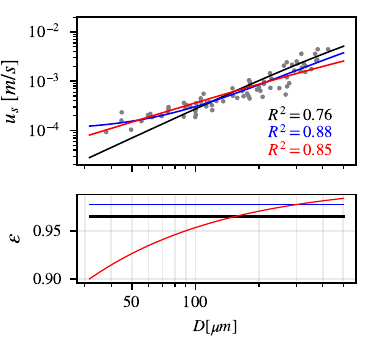}
    \put(-270,220){$a)$}
    \put(-270,90){$b)$}
    \caption{Settling velocity \textit{a)} and porosity \textit{b)} as a function of aggregate size of the of the \textit{FC} (\textcolor{black}{\rule[0.5mm]{0.5cm}{1pt}}), \textit{PE} (\textcolor{blue}{\rule[0.5mm]{0.5cm}{1pt}}), and \textit{FR} (\textcolor{red}{\rule[0.5mm]{0.5cm}{1pt}}) models compared with the experimental data set of \citet{gillard2019}. The experimental data (\textcolor{gray}{$\bullet$}) correspond to the study of \citet{huang1994fractal}. The grey area (\textcolor{mygray}{\rule[0.0mm]{0.5cm}{5pt}}) in (a) represents the normal distribution of the experimental data with a cut-off of the standard deviation $\sigma$.}
    \label{fig:huang}
\end{figure}

\begin{figure}[!ht]
    \centering
    \includegraphics[width = \figwidth]{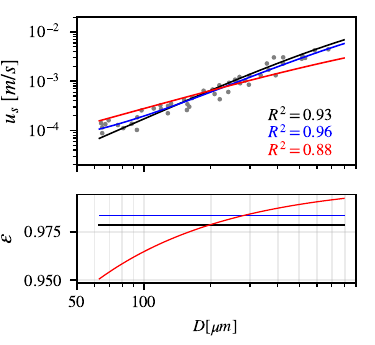}
    \put(-270,220){$a)$}
    \put(-270,90){$b)$}
    \caption{Settling velocity \textit{a)} and porosity \textit{b)} as a function of aggregate size of the of the \textit{FC} (\textcolor{black}{\rule[0.5mm]{0.5cm}{1pt}}), \textit{PE} (\textcolor{blue}{\rule[0.5mm]{0.5cm}{1pt}}), and \textit{FR} (\textcolor{red}{\rule[0.5mm]{0.5cm}{1pt}}) models compared with the experimental data set of \citet{gillard2019}. The experimental data (\textcolor{gray}{$\bullet$}) correspond to the study of \citet{lick1993flocculation}. The grey area (\textcolor{mygray}{\rule[0.0mm]{0.5cm}{5pt}}) in (a) represents the normal distribution of the experimental data with a cut-off of the standard deviation $\sigma$.}
    \label{fig:Lick}
\end{figure}

\begin{figure}[h!]
    \centering
    \includegraphics[width = \figwidth]{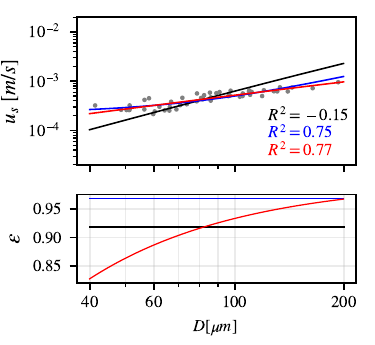}
    \put(-270,220){$a)$}
    \put(-270,90){$b)$}
    \caption{Settling velocity \textit{a)} and porosity \textit{b)} as a function of aggregate size of the of the \textit{FC} (\textcolor{black}{\rule[0.5mm]{0.5cm}{1pt}}), \textit{PE} (\textcolor{blue}{\rule[0.5mm]{0.5cm}{1pt}}), and \textit{FR} (\textcolor{red}{\rule[0.5mm]{0.5cm}{1pt}}) models compared with the experimental data set of \citet{gillard2019}. The experimental data (\textcolor{gray}{$\bullet$}) correspond to the study of \citet{manning1999laboratory}. The grey area (\textcolor{mygray}{\rule[0.0mm]{0.5cm}{5pt}}) in (a) represents the normal distribution of the experimental data with a cut-off of the standard deviation $\sigma$.}
    \label{fig:man_dryer}
\end{figure}

Figures \ref{fig:gibbs}-- \ref{fig:man_dryer} provide a comparison between the aggregate settling velocity measurements reported in \cite{gibbs1985estuarine}, \cite{huang1994fractal}, \cite{lick1993flocculation}, \cite{manning1999laboratory} and various settling velocity models. For the experimental data provided by \cite{gibbs1985estuarine}, \cite{huang1994fractal}, \cite{lick1993flocculation}, we assumed $\rho_s = 2650$ [kg/m$^3$], $\rho_f = 1000$ [kg/m$^3$] and $\nu$ = 9.2 $\times$ 10$^{-7}$ [m$^2$ s$^{-1}$]. For the data of \cite{manning1999laboratory} we assumed $\rho_s = 2265$ [kg/m$^3$], $\rho_f = 1000$ [kg/m$^3$] and $\nu$ = 9.2 $\times$ 10$^{-7}$ [m$^2$ s$^{-1}$]. The fitting parameters for the settling velocity models are similar as reported in the main body of the present paper except for the parameter range for the primary particle size $10^{-6}<D_p<1.5\times 10^{-5}$. Similarly to previous data sets, tables \ref{tab:gibbs}-- \ref{tab:man_dryer} also report the best-fit parameters that correspond to each settling velocity model within a corresponding data set.
\newpage

\begin{table}[h!]
    \centering
    \caption{Best-fit parameters for each settling velocity models provided in figure \ref{fig:ben_0_105}}
    \begin{tabular}{rlccc}
     \multicolumn{2}{c}{Parameters} & $\textit{FC}$ & $\textit{FR}$ & $\textit{PE}$ \\
    \hline
    $\epsilon$ & [-] & 0.9888 & - & 0.9954 \\
    $D_p$ & [$\mu m$] & - & 3.87 & 5.08 \\
    $n_f$ & [-] & - & 2.00 & - \\
    \hline
    \end{tabular}
    \label{tab:ben_0_105}
\end{table}
\newpage
\begin{table}[h!]
    \centering
    \caption{Best-fit parameters for each settling velocity models provided in figure \ref{fig:ben_0_175}}
    \begin{tabular}{rlccc}
     \multicolumn{2}{c}{Parameters} & $\textit{FC}$ & $\textit{FR}$ & $\textit{PE}$ \\
    \hline
    $\epsilon$ & [-] &0.9897 & - & 0.9959 \\
    $D_p$ & [$\mu m$] & - & 4.41 & 4.26 \\
    $n_f$ & [-] & - & 1.92 & - \\
    \hline
    \end{tabular}
    \label{tab:ben_0_175}
\end{table}
\newpage
\begin{table}[h!]
    \centering
    \caption{Best-fit parameters for each settling velocity models provided in figure \ref{fig:ben_2.4_105}}
    \begin{tabular}{rlccc}
     \multicolumn{2}{c}{Parameters} & $\textit{FC}$ & $\textit{FR}$ & $\textit{PE}$ \\
    \hline
    $\epsilon$ & [-] & 0.981 & - & 0.988 \\
    $D_p$ & [$\mu m$] & - & 1.00 & 5.43  \\
    $n_f$ & [-] & - & 2.29 & - \\
    \hline
    \end{tabular}
    \label{tab:ben_2.4_105}
\end{table}
\newpage
\begin{table}[h!]
    \centering
    \caption{Best-fit parameters for each settling velocity models provided in figure \ref{fig:ben_2.4_175}}
    \begin{tabular}{rlccc}
     \multicolumn{2}{c}{Parameters} & $\textit{FC}$ & $\textit{FR}$ & $\textit{PE}$ \\
    \hline
    $\epsilon$ & [-] & 0.981 & - & 0.988 \\
    $D_p$ & [$\mu m$] & - & 1.00 & 5.43  \\
    $n_f$ & [-] & - & 2.29 & - \\
    \hline
    \end{tabular}
    \label{tab:ben_2.4_175}
\end{table}
\newpage
\begin{table}[h!]
    \centering
    \caption{Best-fit parameters for each settling velocity models provided in figure \ref{fig:ben_2.4_500}}
    \begin{tabular}{rlccc}
     \multicolumn{2}{c}{Parameters} & $\textit{FC}$ & $\textit{FR}$ & $\textit{PE}$ \\
    \hline
    $\epsilon$ & [-] & 0.9874 & - & 0.9897 \\
    $D_p$ & [$\mu m$] & - & 1.00 & 2.27  \\
    $n_f$ & [-] & - & 2.2 & - \\
    \hline
    \end{tabular}
    \label{tab:ben_2.4_500}
\end{table}
\newpage
\begin{figure}[!ht]
    \centering
    \includegraphics[width = \figwidth]{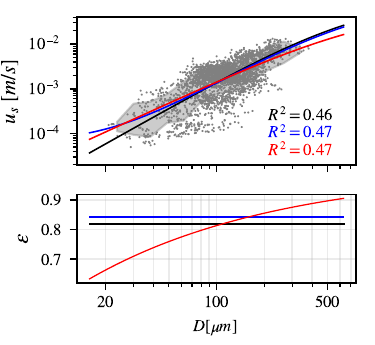}
    \put(-270,220){$a)$}
    \put(-270,90){$b)$}
    \caption{Settling velocity \textit{a)} and porosity \textit{b)} as a function of aggregate size of the of the \textit{FC} (\textcolor{black}{\rule[0.5mm]{0.5cm}{1pt}}), \textit{PE} (\textcolor{blue}{\rule[0.5mm]{0.5cm}{1pt}}), and \textit{FR} (\textcolor{red}{\rule[0.5mm]{0.5cm}{1pt}}) models compared with the experimental data set of \citet{gillard2019}. The experimental data (\textcolor{gray}{$\bullet$}) correspond to Kaolinite sediment from the study of \citet{ye2020oil}. The grey area (\textcolor{mygray}{\rule[0.0mm]{0.5cm}{5pt}}) in (a) represents the normal distribution of the experimental data with a cut-off of the standard deviation $\sigma$.}
    \label{fig:man_kaolinite}
\end{figure}
\begin{table}[h!]
    \centering
    \caption{Best-fit parameters for each settling velocity models provided in figure \ref{fig:man_kaolinite}}
    \begin{tabular}{rlccc}
     \multicolumn{2}{c}{Parameters} & $\textit{FC}$ & $\textit{FR}$ & $\textit{PE}$ \\
    \hline
    $\epsilon$ & [-] & 0.8188 & - & 0.8420 \\
    $D_p$ & [$\mu m$] & - & 1.00 & 1.47  \\
    $n_f$ & [-] & - & 2.64 & - \\
    \hline
    \end{tabular}
    \label{tab:man_kaolinite}
\end{table}
\newpage
\begin{figure}[!ht]
    \centering
    \includegraphics[width = \figwidth]{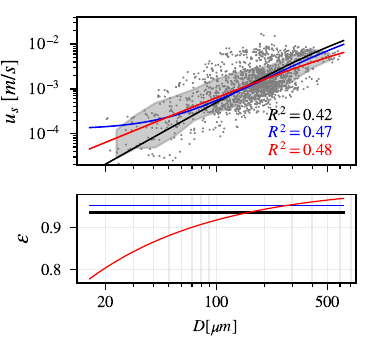}
    \put(-270,220){$a)$}
    \put(-270,90){$b)$}
    \caption{Settling velocity \textit{a)} and porosity \textit{b)} as a function of aggregate size of the of the \textit{FC} (\textcolor{black}{\rule[0.5mm]{0.5cm}{1pt}}), \textit{PE} (\textcolor{blue}{\rule[0.5mm]{0.5cm}{1pt}}), and \textit{FR} (\textcolor{red}{\rule[0.5mm]{0.5cm}{1pt}}) models compared with the experimental data set of \citet{gillard2019}. The experimental data (\textcolor{gray}{$\bullet$}) correspond to Bentonite sediment from the study of \citet{ye2020oil}. The grey area (\textcolor{mygray}{\rule[0.0mm]{0.5cm}{5pt}}) in (a) represents the normal distribution of the experimental data with a cut-off of the standard deviation $\sigma$.}
    \label{fig:man_bentonite}
\end{figure}
\begin{table}[h!]
    \centering
    \caption{Best-fit parameters for each settling velocity models provided in figure \ref{fig:man_bentonite}}
    \begin{tabular}{rlccc}
     \multicolumn{2}{c}{Parameters} & $\textit{FC}$ & $\textit{FR}$ & $\textit{PE}$ \\
    \hline
    $\epsilon$ & [-] & 0.936 & - & 0.952 \\
    $D_p$ & [$\mu m$] & - & 1.00 & 9.24  \\
    $n_f$ & [-] & - & 2.46 & - \\
    \hline
    \end{tabular}
    \label{tab:man_bentonite}
\end{table}
\newpage
\begin{figure}[!ht]
    \centering
    \includegraphics[width = \figwidth]{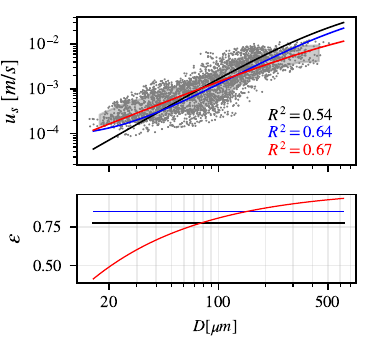}
    \put(-270,220){$a)$}
    \put(-270,90){$b)$}
    \caption{Settling velocity \textit{a)} and porosity \textit{b)} as a function of aggregate size of the of the \textit{FC} (\textcolor{black}{\rule[0.5mm]{0.5cm}{1pt}}), \textit{PE} (\textcolor{blue}{\rule[0.5mm]{0.5cm}{1pt}}), and \textit{FR} (\textcolor{red}{\rule[0.5mm]{0.5cm}{1pt}}) models compared with the experimental data set of \citet{gillard2019}. The experimental data (\textcolor{gray}{$\bullet$}) correspond to mixed sediment of Bentonite and Kaolinite from the study of \citet{ye2020oil}. The grey area (\textcolor{mygray}{\rule[0.0mm]{0.5cm}{5pt}}) in (a) represents the normal distribution of the experimental data with a cut-off of the standard deviation $\sigma$.}
    
    \label{fig:man_kaol_bent}
\end{figure}
\begin{table}[h!]
    \centering
    \caption{Best-fit parameters for each settling velocity models provided in figure \ref{fig:man_kaol_bent}}
    \begin{tabular}{rlccc}
     \multicolumn{2}{c}{Parameters} & $\textit{FC}$ & $\textit{FR}$ & $\textit{PE}$ \\
    \hline
    $\epsilon$ & [-] & 0.7789 & - & 0.8549 \\
    $D_p$ & [$\mu m$] & - & 6.63 & 1.5  \\
    $n_f$ & [-] & - & 2.39 & - \\
    \hline
    \end{tabular}
    \label{tab:man_kaol_bent}
\end{table}
\newpage
\begin{table}[h!]
    \centering
    \caption{Best-fit parameters for each settling velocity models provided in figure \ref{fig:gibbs}}
     \begin{tabular}{rlccc}
     \multicolumn{2}{c}{Parameters} & $\textit{FC}$ & $\textit{FR}$ & $\textit{PE}$ \\
    \hline
    $\epsilon$ & [-] & 0.8484 & - & 0.9620 \\
    $D_p$ & [$\mu m$] & - & 11.1 & 11  \\
    $n_f$ & [-] & - & 1.9 & - \\
    \hline
    \end{tabular}
    \label{tab:gibbs}
\end{table}
\newpage
\begin{table}[h!]
    \centering
    \caption{Best-fit parameters for each settling velocity models provided in figure \ref{fig:huang}}
     \begin{tabular}{rlccc}
     \multicolumn{2}{c}{Parameters} & $\textit{FC}$ & $\textit{FR}$ & $\textit{PE}$ \\
    \hline
    $\epsilon$ & [-] & 0.9654 & - & 0.9775 \\
    $D_p$ & [$\mu m$] & - & 1.0 & 5.49  \\
    $n_f$ & [-] & - & 2.33 & - \\
    \hline
    \end{tabular}
    \label{tab:huang}
\end{table}

\newpage
\begin{table}[h!]
    \centering
    \caption{Best-fit parameters for each settling velocity models provided in figure \ref{fig:Lick}}
     \begin{tabular}{rlccc}
     \multicolumn{2}{c}{Parameters} & $\textit{FC}$ & $\textit{FR}$ & $\textit{PE}$ \\
    \hline
    $\epsilon$ & [-] & 0.9785 & - & 0.9832 \\
    $D_p$ & [$\mu m$] & - & 1.0 & 3.09  \\
    $n_f$ & [-] & - & 2.27 & - \\
    \hline
    \end{tabular}
    \label{tab:Lick}
\end{table}
\newpage
\begin{table}[h!]
    \centering
    \caption{Best-fit parameters for each settling velocity models provided in figure \ref{fig:man_dryer}}
     \begin{tabular}{rlccc}
     \multicolumn{2}{c}{Parameters} & $\textit{FC}$ & $\textit{FR}$ & $\textit{PE}$ \\
    \hline
    $\epsilon$ & [-] & 0.919 & - & 0.9687 \\
    $D_p$ & [$\mu m$] & - & 7.3 & 9.71  \\
    $n_f$ & [-] & - & 1.97 & - \\
    \hline
    \end{tabular}
    \label{tab:man_dryer}
\end{table}

\newpage
\textbf{Declaration of generative AI and AI-assisted technologies in the writing process}

During the preparation of this work the authors used \textit{deepl.com} in order to improve language and readability. After using this tool/service, the author(s) reviewed and edited the content as needed and take(s) full responsibility for the content of the publication.
\newpage

\bibliographystyle{elsarticle} 
\bibliography{elsarticle}

\end{document}